\documentclass[a4paper,UKenglish,numberwithinsect,cleveref, autoref, thm-restate]{lipics-v2021}
\usepackage{arxiv-lipics-v2021}
\nolinenumbers

\usepackage{xspace}
\usepackage{bm}
\usepackage[colorinlistoftodos]{todonotes}
\usepackage{tikz}
\usepackage{tabularx}
\usepackage{environ}
\usetikzlibrary{arrows.meta, positioning}
\usepackage{amsmath}
\usepackage{amssymb}
\usepackage{newfile}
\usepackage{totcount}

\usetikzlibrary{automata}
\usetikzlibrary{positioning}
\usetikzlibrary{shapes}
\usetikzlibrary{arrows}
\usetikzlibrary{mindmap,trees}
\usetikzlibrary{decorations.pathreplacing}
\usetikzlibrary{arrows.meta}
\usetikzlibrary{calc}
\usetikzlibrary{backgrounds}
\usetikzlibrary{shapes.misc}
\usetikzlibrary{svg.path}
\usetikzlibrary{tikzmark}

\title{Regular Separability for Bidirected VASS} %

\author{Łukasz Orlikowski}
{Max Planck Institute for Software Systems, Kaiserslautern, Germany}
{lorlikowski@mpi-sws.org}
{https://orcid.org/0009-0001-4727-2068}
{}
\author{Georg Zetzsche}
{Max Planck Institute for Software Systems, Kaiserslautern, Germany}
{georg@mpi-sws.org}
{https://orcid.org/0000-0002-6421-4388}
{}
\authorrunning{\L. Orlikowski and G. Zetzsche} %

\funding{\flag[3cm]{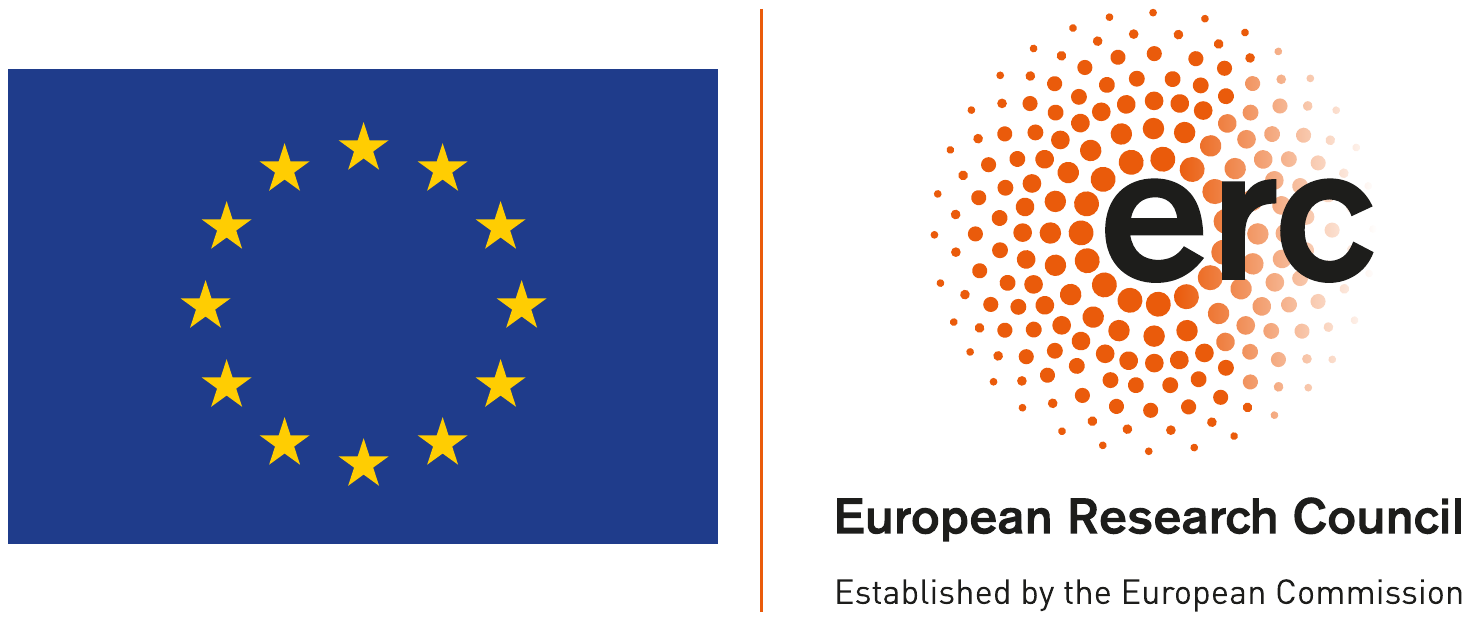}Funded by the European Union (ERC, FINABIS, 101077902). Views and opinions expressed are however those of the author(s) only and do not necessarily reflect those of the European Union or the European Research Council Executive Agency. Neither the European Union nor the granting authority can be held responsible for them.
	In addition, the first author was supported by the European Research Council (ERC) grant POLARIS, agreement no. 101231447.
}

\Copyright{Łukasz Orlikowski and Georg Zetzsche} %



\category{} %

\acknowledgements{We are grateful to Wojciech Czerwinski for the idea to study regular separablity for bidirected systems.}%

\crefname{appsec}{Appendix}{Appendices}

\tikzset{gadget/.style={->,>=stealth,initial text=,auto,on grid,scale=1,inner sep=1pt}}

\tikzset{every state/.style={minimum size=10pt, inner sep=1pt,fill=black!10,draw=black!70,thick}}

\begin{document}
	\newcommand{\regsep}{\textup{Regular separability}\xspace}
\newcommand{\regsepop}{\mathbin{|}}
\newcommand{\regseparable}{regular separable}
\newcommand{\ackermann}{\ensuremath{\mathsf{ACKERMANN}}\xspace}
\newcommand{\ackermanncomplete}{\ackermann-complete\xspace}
\newcommand{\nlcomplete}{\ensuremath{\mathsf{NL}}-complete\xspace}
\newcommand{\expspace}{\ensuremath{\mathsf{EXPSPACE}}\xspace}
\newcommand{\expspacecomplete}{\expspace-complete\xspace}
\newcommand{\expspacehard}{\expspace-hard\xspace}
\newcommand{\conp}{\mathsf{coNP}\xspace}
\newcommand{\zt}{$\textup{VASS}_z$\xspace}
\newcommand{\zerotest}[1]{$#1\textup{-VASS}_z$\xspace}
\newcommand{\integerzerotest}{\Z\zerotest{}}
\newcommand{\fsep}{$\mathcal{F}$-Sep\xspace}
\newcommand{\fseparable}{$\mathcal{F}$-separable\xspace}
\newcommand{\set}[1]{\{#1\}}
\newcommand{\setof}[2]{\set{#1 \mid #2}}
\newcommand{\Z}{\mathbb{Z}}
\newcommand{\N}{\mathbb{N}}
\newcommand{\countletters}[2]{\#_{#1}(#2)}
\newcommand{\zvass}{$\Z$\textup{-VASS}\xspace}
\newcommand{\ptime}{\ensuremath{\mathsf{P}}}
\newcommand{\gztodo}[1]{\todo[inline, linecolor=blue, backgroundcolor=blue!30, bordercolor=black]{\textbf{Georg:} #1}}
\newcommand{\lotodo}[1]{\todo[inline, linecolor=yellow, backgroundcolor=yellow!30, bordercolor=black]{\textbf{{\L}ukasz:} #1}}
\newcommand{\twocm}{$2$-counter machine\xspace}
\newcommand{\trans}[2]{$#1 \rightarrow^* #2$}
\newcommand{\prefix}[2]{\textup{prefix}_#1(#2)}
\newcommand{\minprefix}[2]{\min_{1 \leq i \leq |#2|}#1(\prefix{i}{#2})}
\newcommand{\maxprefix}[2]{\max_{1 \leq i \leq |#2|}#1(\prefix{i}{#2})}
\newcommand{\mywlog}{w.l.o.g.\xspace}
\newcommand{\nerode}[1]{u \equiv_{#1} v \iff \forall_{w \in \Sigma^*} (uw \in #1 \iff vw \in #1)}
\newcommand{\eff}{\textup{eff}}
\newcommand{\rev}{\textup{rev}}
\newcommand{\cover}{\ensuremath{\ge}}
\newcommand{\Wlog}{W.l.o.g\xspace}
\newcommand{\size}{\textup{size}}
\newcommand{\norm}{\textup{norm}}
\renewcommand{\vec}[1]{\myvec{#1}}
\newcommand{\myvec}[1]{\bm{#1}}
\newcommand{\supp}{\textup{supp}}
\makeatletter
\newcommand{\problemtitle}[1]{\gdef\@problemtitle{#1}}%
\newcommand{\probleminput}[1]{\gdef\@probleminput{#1}}%
\newcommand{\problemquestion}[1]{\gdef\@problemquestion{#1}}%
\NewEnviron{problem}{
  \probleminput{}\problemquestion{}%
  \BODY%
  \par\addvspace{.5\baselineskip}
  \noindent
  \begin{tabularx}{\textwidth}{@{\hspace{\parindent}} l X c}
    \textbf{\textup{Input:}} & \@probleminput \\%
    \textbf{\textup{Question:}} & \@problemquestion%
  \end{tabularx}
  \par\addvspace{.5\baselineskip}\noindent
}
\makeatother
\newcommand{\skel}{\textup{skel}}
\newcommand{\1}{\mathbf{1}}
\newcommand{\reach}{\textup{REACH}}
\newcommand{\myparagraph}[1]{\vskip 0.1 cm \noindent \textbf{#1.}}

	\maketitle

	\begin{abstract}
		The regular separability problem asks, given two languages, whether there exists a regular language that contains one of them and is disjoint from the other. We study regular separability for bidirected
		VASS. We first show that the problem remains Ackermann-complete when the input consists of two languages recognised by bidrected VASS. To find a subclass of VASS in which the problem has lower complexity we define alphabetically bidirected VASS as ones
		where bidirectedness is also visible in the input letters. Our main result establishes that regular separability for alphabetically bidirected VASS is \expspace-complete. The key to this result is that, in this setting, regular separability coincides with disjointness.
		
		As a first ingredient in this equivalence, we also show that for alphabetically
		bidirected $\Z$-VASS, regular
		separability and disjointness are equivalent. Consequently, regular separability for these systems is in polynomial-time. In fact, it is log-space inter-reducible with non-solvability of integer linear
		Diophantine equation systems.
		
		Finally, we observe that the picture changes dramatically in the presence
		of a zero test, namely, we show that regular separability for
		alphabetically bidirected one-counter automata is undecidable.
	\end{abstract}
	\vspace{1em}
	\newline
	\textcolor{lipicsGray}{\fontsize{9}{12}\sffamily\bfseries
		AI Disclosure\enskip}No AI tools were used in this work.
	\newpage
	\section{Introduction}
\emph{Regular separability} is a decision problem for two given languages
$L_1,L_2\subseteq\Sigma^*$, and it asks whether they are \emph{regularly
separable}, i.e.\ whether there is a regular language $S\subseteq\Sigma^*$ such
that $L_1\subseteq S$ and $L_2\cap S=\emptyset$. Regular separability has
received a significant amount of attention. After some discouraging
undecidability results on context-free languages in the
1970s~\cite{SzymanskiW-sicomp76,DBLP:journals/jacm/Hunt82a}, regular
separability has been studied intensively over the last decade for \emph{vector
addition systems with states} (VASS), which have finite many control states
with access to counters that can be incremented and decremented, but not
zero-tested. Moreover, the counters have to stay negative during the run.
Regular separability has been studied for integer
VASS~\cite{z_vass_separability_classical,new_zvass}, combination of VASS and
subclasses~\cite{DBLP:conf/lics/CzerwinskiZ20}, B\"{u}chi
VASS~\cite{DBLP:conf/icalp/0001K0Z24}, and VASS coverability languages (more
generally, well-structured transition
systems)~\cite{CzerwinskiLMMKS18,regular}, VASS reachability sets~\cite{DBLP:conf/stacs/ClementeCLP17}, and one-dimensional VASS~\cite{ocn_separability}. This culminated in the celebrated
decidability result for regular separability of general
VASS~\cite{DBLP:conf/lics/Keskin024}.

\subparagraph*{Remaining challenges} Currently, three challenges remain in
connection with regular separability. First, decidability is not known in the
``CFL vs.\ VASS'' setting, i.e.\ when one input language is context-free and
the other is a VASS (reachability) language. Note that this significantly
generalizes the ``VASS vs.\ VASS'' setting, because any two VASS languages
$L_1,L_2\subseteq\Sigma^*$ are regularly separable if and only if the (linear)
context-free language $E=\{w\#w^{\rev} \mid w\in\Sigma^*\}$ is regularly separable
from the VASS language $K=L_1\#L_2^{\rev}$ (in each case, $\cdot^{\rev}$
denotes string reversal). 

The second challenge concerns the mysterious relationship between regular
separability and the \emph{intersection problem}, which asks whether two given
languages intersect. As of now, almost all existing results on decidability and
complexity of regular separability are in agreement with the corresponding
intersection problem: In other words, in almost all settings, deciding regular
separability of $L_1,L_2$ has the same complexity as deciding whether $L_1\cap
L_2=\emptyset$. The only exceptions to this are artificially constructed
language classes that show that each problem can be decidable without the
other~\cite[Thms.~4.1\&5.2]{DBLP:conf/fsttcs/ThinniyamZ19}. However, for all
natural classes of infinite-state systems, the two problems agree in their
decidability and complexity. In fact, under mild assumptions, if $L_1,L_2$ are
accepted by well-structured transition systems (WSTS), then regular
separability is even equivalent to disjointness~\cite{CzerwinskiLMMKS18}. This
close agreement across infinite-state systems raises the question of whether
there is a deeper connection, perhaps even a reduction technique, between
regular separability and intersection. Note that the aforementioned setting
``CFL vs.\ VASS'' is a special case of this: It is known the intersection
between CFL and VASS languages is decidable (it is equivalent to reachability
in pushdown VASS~\cite{DBLP:conf/lics/GuttenbergKM26}), but regular separability
remains unknown.

The third challenge is to identify subclasses of VASS where the complexity of
regular separability is lower than the general Ackermann-complete
setting~\cite[Thm.~3.1]{DBLP:conf/lics/Keskin024}. 

\subparagraph*{Bidirected VASS} In this work, we make progress on the second
and third challenge above, by studying regular separability for the subclass of
\emph{bidirected} VASS. These are VASS where for every transition, there is
another transition with opposite effect. Bidirectedness is a restriction that
has been studied across recent decades for various types of infinite-state
systems. This is because some types of static analysis directly correspond to
reachability in bidirected (counter or pushdown)
systems~\cite{DBLP:conf/slp/Reps97}, but also because bidirectedness enables
efficient algorithms for pushdown
systems~\cite{DBLP:journals/pacmpl/ChatterjeeCP18},
VASS~\cite{mayr1982complexity,DBLP:journals/iandc/KoppenhagenM00,DBLP:journals/pacmpl/KjelstromP22,DBLP:journals/pacmpl/KrishnaLPT24},
pushdown VASS~\cite{DBLP:conf/icalp/GanardiMPSZ22} and other kinds of
systems~\cite{DBLP:conf/lics/GanardiMZ22} (see
\cite{DBLP:journals/siglog/Pavlogiannis22} for an overview).

\subparagraph*{Contributions} We study regular separability for bidirected
VASS. We first note that it is not obvious to find the right definition: If we
only assume that we are given bidirected VASS for input languages $L_1$ and $L_2$, then we
show that it is still Ackermann-complete to decide whether $L_1$ and $L_2$ are
regularly separable. However, a slightly different perspective will indeed
yield lower complexity. We define \emph{alphabetically bidirected VASS} as ones
where bidirectedness is also visible in the input letters: The input alphabet
$\Sigma$ consists of two parts $\Sigma_1$ and $\bar{\Sigma}_1$, where
$\bar{\Sigma}_1=\{\bar{a}\mid a\in\Sigma_1\}$ are considered opposite to
$\Sigma_1$. And crucially, the opposite of a transition also carries the
opposite label. 

For alphabetically bidirected VASS, we show that the complexity of regular
separability indeed drops significantly. Our \emph{main result} is that for
alphabetically bidirected VASS, regular separability is \expspace-complete. As
mentioned above, this contrasts with Ackermann-completeness for general VASS.
To this end, we show that for alphabetically bidirected VASS, regular
separability coincides with disjointness.

As a first ingredient in this equivalence, we also show that for alphabetically
bidirected $\Z$-VASS (i,e.\ VASS where counters can drop below zero), regular
separability and disjointness coincide. This yields a result of independent
interest: Regular separability for alphabetically bidirected $\Z$-VASS is in
polynomial time (in fact, inter-reducible with non-solvability of integer linear
Diophantine equation systems). Again, this is in contrast to
$\conp$-completeness for general $\Z$-VASS~\cite[Thm.~3.6]{new_zvass}.

However, in our \emph{second main result}, we show that for one-counter
automata (i.e.\ VASS with one zero-testable counter), regular separability in
the alphabetically bidirected setting is undecidable, just like for
unrestricted one-counter automata~\cite[Thm.~2]{ocn_separability}.

\subparagraph{Lower complexity and connection to disjointness} This means, with
the alphabetically bidirected setting, we identify two subclasses of VASS with
lower complexity of regular separability, and thus contribute to the
\emph{third challenge} above. In addition, we make progress on the \emph{second
challenge}: This is because our algorithms for VASS and $\Z$-VASS rely on the
equivalence between regular separability and disjointness. In other words, they
enjoy the same property as most languages of WSTS~\cite{CzerwinskiLMMKS18}.

\subparagraph*{Key insights} Our proofs rely on several new insights. As
mentioned above, the first ingredient is that in alphabetically bidirected
$\Z$-VASS, regular separability and disjointness are equivalent. To this end,
we work with a characterization \cite[Thm.~4.6]{new_zvass} of regular
inseparability of $\Z$-VASS in terms of solvability of a linear equation system
associated with the $\Z$-VASS. These solutions correspond to \emph{integer}
linear combinations of cycles, i.e.\ where some cycles can occur with negative
multiplicity. Roughly speaking, we observe that in bidirected systems, such
solutions can be converted into proper runs, by replacing negative-multiplicity
cycles $\pi$ with positive multiples of their opposite cycle $\bar{\pi}$.

Then for (non-integer) VASS, our proof consists of several steps. First, we
consider the case that all counters of the input VASS are simultaneously
unbounded (i.e.~in a product of the input VASS, there are runs that make all
counters arbitrarily large at the same time). In this case, we consider three
overapproximations: (i)~the forward coverability language, (ii)~the backward
coverability language, and (iii)~the words accepted in $\Z$-semantics (i.e.\
where counter can drop below zero). By relying on the equivalence of regular
separability and disjointness for coverability VASS~\cite{regular} (in the case
of (i)\&(ii)) and alphabetically bidirected $\Z$-VASS (for (iii)), we then know
that if any of those overapproximations yield empty intersection, we have
regular separability. However, if each pair of overapproximations intersects,
simultaneous unboundedness allows us to construct a word in the intersection.

If the counters are not all simultaneously unbounded, we rely on a result about
bidirected VASS~\cite[Lemma 6.2]{DBLP:conf/lics/GanardiMZ22}, which says that
all unbounded counters in a bidirected VASS are even simultaneously unbounded.
This means, there must be one counter that is bounded in the product
construction. Roughly speaking, this allows us to eliminate that counter and
argue by induction on the total number of counters.

In our undecidability proof for alphabetically bidirected OCA, we follow the
same strategy as in the undecidability for general OCA~\cite{ocn_separability}.
However, the latter crucially relies on the fact that in a two-counter machine
that terminates after $t$ steps, if $R_i$ denotes the set of runs ignoring
counter $i$, then $R_1$ and $R_2$ can agree on a prefix of length at most $t$.
However, if we attempt to translate this into a bidirected setting by adding
backward edges, then we introduce arbitrary-length common prefixes. Therefore,
we rely on a result of Morita~\cite{reversible_automata} that shows
Turing-completeness of \emph{reversible} (i.e.\ backward deterministic)
two-counter machines. We observe that here, even with added backward edges, a
regular language can differentiate prefixes of $R_1$ from $R_2$, even though
they may agree on arbitrary-length prefixes.

\subparagraph*{Structure of the paper} After introducing basic notions in
\cref{sec:prelim}, we discuss the aforementioned naive notion of regular
separability in bidirected VASS in \cref{sec:attempt}. In
\cref{sec:zvass}, we prove the equivalence between regular separability and
disjointness in alphabetically bidirected \zvass. \Cref{sec:vass} then shows
out main result: equivalence of regular separability and disjointness in
alphabetically bidirected VASS. Finally, \cref{sec:undecidability} proves undecidability
of regular separability of alphabetically bidirected OCA.

	\section{Preliminaries}\label{sec:prelim}
\subparagraph*{Basic notions}
For $a,b \in \N$ such that $a \leq b$ we write $[a, b]$ for the set of integers $\{a, a+1, \ldots, b\}$.
For $a \in \N$ we write $[a]$ for the set $[1, a]$. For $a \in \Z$ we write $\vec{a}$ for the constant vector $(a, \ldots, a) \in \Z^d$.
For a vector $\myvec v \in \mathbb{Z}^d$, we denote its $i$-th component by $\myvec v_i$ and the vector obtained by removing its $i$-th component by $\pi_{-i}(\myvec v)$. We also define the support of the vector $v \in \Z^d$ as the set of all non-zero components, i.e., $\supp(\myvec v)= \{i \in [1,d] \mid \myvec v_i \neq 0\}$. For two vectors $\myvec u, \myvec v \in \N^d$, we write $\myvec u \leq \myvec v$ if for all $i \in [d]$ we have $\myvec u_i \leq \myvec v_i$. For $i \in \{1,\ldots,d\}$, let $\myvec e_i \in \mathbb{Z}^d$ denote the $i$-th standard basis vector, whose $i$-th component is $1$ and all other components are $0$. For a word $w$ and a letter $a$, we denote by $\#_a(w)$ the number of letters $a$. Similarly, for a word $w$ and a set of letters $\Sigma$, we denote by $\#_\Sigma(w)$ the number of letters from $\Sigma$ in $w$. For a set of letters $\Sigma$ we denote by $\Sigma_{\varepsilon}$ the set $\Sigma \cup \{\varepsilon\}$. For a word $w$ and $i \in [|w|]$ we denote by $\prefix{i}{w}$ the prefix of $w$ of length $i$.

\subparagraph*{Vector Addition Systems with States}
A $d$-dimensional Vector Addition System with States ($d$-VASS) is a nondeterministic finite automaton with $d$ non-negative integer counters.  Formally, we define VASS $V$ as  $V = (\Sigma, Q, \delta, q_i, q_f)$ where $\Sigma$ is a finite alphabet, $Q$ is a finite set of automaton states, $\delta \subseteq Q \times \Sigma_\varepsilon \times \Z^d \times Q$ is a transition relation, $q_i$ is an initial state and $q_f$ is the final state. For a transition $t = (s, a, \myvec v, s') \in \delta$, we say that the transition is over the letter $a$ or that the transition reads the letter $a$. We also write $\eff(t)$ for the effect of the transition $t$, which is $\myvec v$. For a VASS $V = (\Sigma, Q, \delta, q_i, q_f)$ we denote by $\rev(V)$ the VASS $(\Sigma, Q, \delta, q_f, q_i)$

A $d$-VASS can be seen as an infinite-state labelled transition system in which each configuration is a pair $(s, \myvec u) \in Q \times \N^d$. We denote such a configuration as $s(\myvec u)$. A transition $t = (s, a, \myvec v, s') \in \delta$ can be fired in a configuration $q(\myvec u)$ if, and only if $q=s$ and $\myvec u' \geq 0$ where $\myvec u'=\myvec u + \myvec v$. After firing the transition, the configuration is changed to $s'(\myvec u')$. Whenever a VASS is part of the input we assume that numbers on transitions are encoded in binary. We also define a run as a sequence of transitions, which can be fired one after another from some configuration. For a run $\rho=t_1t_2\ldots t_n$ we write $\eff(\rho)$ for $\Sigma_{i=1}^n \eff(t_i)$. If we want to say something only about the $ j$-th (for $j \in [d]$) entry of the $\eff(\rho)$, we write $\eff_j(\rho)$, and for the effect on the subset of counters $I$ we write $\eff_I(\rho)$. We say that a run $\rho$ is from configuration $q(\myvec u)$ to $p(\myvec u')$ if the sequence of transitions can be fired from $q(\myvec u)$ and the final configuration is $p(\myvec u')$. We say that a run is a cycle if the state of its initial configuration is the same as the state of the final configuration. For two runs $\rho_1$, $\rho_2$ we denote by $\rho_1\rho_2$ the run obtained by concatenation of $\rho_1$ and $\rho_2$. We say that a run $\rho$ is over (or reads) a word 
$w \in \Sigma^*$ if and only if the sequence of all non-$\varepsilon$ labels of transitions of $\rho$ is exactly $w$ (preserving the order of transitions). We say that the length of a run $\rho$ is equal to $n$ if it consists of $n$ transitions. For the length of a run, we write $|\rho|$. For a VASS $V$, configuration $s \in Q \times \N^d$, and state $q$, we denote by $\reach(V, s, q)$ the set of vectors such that if $\myvec v \in \reach(V, s, q)$ then there is a run from the configuration $s$ to configuration $q(\myvec v)$ in $V$. 

We consider a variant of VASS, called integer VASS (\zvass), where the nonnegativeness constraint is dropped. 
Syntactically, \zvass is the same as VASS, namely it consists of a finite alphabet, a finite set of states, a finite set of transitions, an initial state $q_i$, and a final state $q_f$. Semantically, configurations of a \zvass are $Q \times \Z^d$ (instead of $Q \times \N^d$), while all definitions (run, reachability set, etc.) are the same as in case of VASS, except the fact that a transition $t = (s, a, \myvec v, s')$ can be fired in a configuration $q(\myvec u)$ if, and only if, $q = s$ (we do not restrict that the counters cannot drop below zero). Equivalently, we may also speak of
$\Z$-configurations and $\Z$-runs of a VASS, i.e., configurations and runs where the nonnegativeness constraint is dropped. 

We also consider one counter automaton (OCA), which is a $1$-VASS, where the counter can be additionally zero-tested. Syntactically and semantically, OCA is almost the same as $1$-VASS, namely, it consists of a finite alphabet, a finite set of states, a finite set of transitions, an initial state $q_i$, and a final state $q_f$. The only difference is that it can have special transitions, which can be fired only if the counter is equal to zero. 

\subparagraph*{Bidirected VASS}
We say that a VASS $V$ is bidirected if the existence of a transition $t = (p, a, \myvec v, q) \in \delta$ implies that there also exists a transition $t' = (q, b, -\myvec v, p) \in \delta$ for some $b \in \Sigma$. We call $t'$ a reverse transition of $t$ and denote it as $\overline{t}$. Moreover, we say that $V$ is alphabetically bidirected if (i) its alphabet $\Sigma = \Sigma_1 \cup \overline{\Sigma_1}$ where $\overline{\Sigma_1} = \setof{\overline{a}}{a \in \Sigma_1}$ (ii) the existence of a transition $t = (p, a, \myvec v, q) \in \delta$ implies that there also exists a transition $t' = (q, \overline{a}, -\myvec v, p) \in \delta$ where we set $\overline{\overline{a}} = a$. We call $t'$ the reverse transition of $t$ and denote it as $\overline{t}$. Notice that in the case of alphabetically bidirected VASS, the reverse transition is uniquely determined. For a word $w=a_1a_2\ldots a_n \in \Sigma^*$ we define its reverse $\overline{w}$ as $\overline{a_n}\overline{a_{n-1}}\ldots \overline{a_1}$. Similarly for a run $\rho=t_1t_2\ldots t_n$ we define its reverse $\overline{\rho}$ as $\overline{t_n}\overline{t_{n-1}}\ldots\overline{t_1}$. We extend the notion of bidirected and alphabetically bidirected to \zvass and OCA, noting that for OCA, if a transition zero-tests the counter then the reverse transition also needs to zero-test the counter. 

\subparagraph*{Languages of VASSs}
A run of a VASS is accepting if it starts in the initial configuration $q_i(\vec{0})$ and ends in the accepting configuration $q_f(\vec{0})$. 
For a VASS $V$, we define its language as the set of all words read by accepting runs, and we denote it by $L(V)$. Languages defined in this way are called reachability languages. Sometimes we consider languages with other acceptance conditions (coverability languages); in that case, we indicate it clearly. The coverability language of a VASS is defined as the set of all words read by runs from $q_i(\vec{0})$ to $q_f(\vec{v})$ for some $\vec{v} \in \N^d$. Therefore, it is a superset of the reachability language. We denote it by $L_\cover(V)$. Sometimes we consider the language of a VASS when it is interpreted as a \zvass, namely, we allow the counters to have negative values along the run. We denote the language of a \zvass $V$ by $L_{\Z}(V)$, or simply by $L(V)$ when it is clear from the context that $V$ is a \zvass. It consists of all words read along $\Z$-runs from $q_i(\vec{0})$ to $q_f(\vec{0})$. Similarly, we define $L(V)$ if $V$ is an OCA instead of a standard VASS.

\subparagraph*{Regular separability}
We call two languages $K, L \subseteq \Sigma^*$ regularly separable if there exists a regular language $R \subseteq \Sigma^*$ such that $K \subseteq R$ 
and $L \cap R = \emptyset$. In that case, we write $K \regsepop L$ and call $R$ a regular separator of $K$ and $L$. We are mainly interested in the regular separability problem, namely, given two languages, decide whether there exists a regular separator that separates them. Sometimes we analyse similar problems for different classes of separators than regular languages. Concretely, let $\mathcal{F}$ be a class of languages. Then the \fsep problem asks whether for languages $K$ and $L$ there exists a language $R \in \mathcal{F}$ such that $K \subseteq R$ and $L \cap R = \emptyset$. 

\subparagraph*{Recognizable separability}
A subset $L \subseteq M$ of a monoid $M$ is recognizable if there is a morphism
$\phi: M \rightarrow F$ into some finite monoid $F$ such that $\phi^{-1}(\phi(L)) = L$. We say that sets $K, L \subseteq M$ are
recognizably separable if and only if there exists a recognizable $S \subseteq M$ such that $K \subseteq S$ and $S \cap L = \emptyset$. Similarly, as in the case of regular separability, if sets $K$ and $L$ are recognizably separable, then we write $K \regsepop L$.

	\section{A naive attempt: non-alphabetic bidirectedness}\label{sec:attempt}
In this section, we explain why bidirectedness alone is not sufficient for obtaining better complexity bounds and why we need to consider alphabetically bidirected VASSs.
\begin{lemma}\label{lem:bidirected}
    For any class of languages $\mathcal{F}$ closed under intersection and union with piecewise testable languages\footnote{Recall that a regular language is piecewise testable if it is a finite boolean combination of languages of the form $\Sigma^*a_1\Sigma^*a_2\ldots\Sigma^*a_n\Sigma^*$, where $a_i \in \Sigma$ for $i \in [n]$.} \fsep of two VASSs languages can be reduced to \fsep of two bidirected VASSs languages.
\end{lemma}
\begin{proof}
    Let us fix two VASSs $V_1$ and $V_2$. We will constuct two bidirected VASSs $V_1'$ and $V_2'$ such that $L(V_1)$ and $L(V_2)$ are \fseparable if and only $L(V_1')$ and $L(V_2')$ are \fseparable. 
    
    Let $\Sigma$ be the set of all the letters used by $V_1$ and $V_2$. Let us take two letters $a_1, a_2 \notin \Sigma$. For $i \in [1,2]$, define $V_i'$ as a copy of $V_i$ with additional transitions, namely, for each transition $t=(p,b,\myvec v, q)$ of $V_i$, we construct transition $t'=(q,a_i,-\myvec v, p)$ in $V_i'$. Observe that clearly from the construction, both $V_1'$ and $V_2'$ are bidirected.

    Firstly, we show that if $L(V_1)$ and $L(V_2)$ are \fseparable then $L(V_1')$ and $L(V_2')$ are \fseparable. Let $L \in \mathcal{F}$ such that $L(V_1) \subseteq L$ and $L(V_2) \cap L = \emptyset$. As $L(V_1)$ and $L(V_2)$ are over $\Sigma$ we can assume $L$ is over $\Sigma$ (otherwise take $L \cap \Sigma^* \in \mathcal{F}$ as a separator). Let $L' = L \cup \Sigma'^*a_1\Sigma'^* \in \mathcal{F}$ where $\Sigma' = \Sigma \cup \{a_1\}$. Clearly $L(V_1') \subseteq L'$. Hence it enough to show that $L' \cap L(V_2') = \emptyset$. Assume, towards contradiction, that there exists $w \in L' \cap L(V_2')$. Observe, that $L'$ is a language over $\Sigma \cup \{a_1\}$ and $L(V_2')$ is a language over $\Sigma \cup \{a_2\}$. Hence $w \in \Sigma^*$. Observe, that $L' \cap \Sigma^* = L$ and $L(V_2') \cap \Sigma^* = L(V_2)$. Thus, we can observe that $w \in L(V_2)$ and $w \in L$, which contradicts the fact that $L \cap L(V_2) = \emptyset$. Hence $L' \cap L(V_2') = \emptyset$ and $L(V_1')$ and $L(V_2')$ are \fseparable.

    Now, we show that if $L(V_1')$ and $L(V_2')$ are \fseparable then $L(V_1)$ and $L(V_2)$ are \fseparable. Let $L \in \mathcal{F}$ such that $L(V_1') \subseteq L$ and $L(V_2') \cap L = \emptyset$. Let $L' = L \cap \Sigma^* \in \mathcal{F}$. Clearly $L' \cap L(V_2') = \emptyset$. Moreover, observe, that $L(V_1) = L(V_1') \cap \Sigma^* \subseteq L \cap \Sigma^* = L'$.
\end{proof}
Class of regular languages is closed under intersection and union with piecewise testable languages. Therefore, because regular separability is \ackermanncomplete for VASSs \cite{DBLP:conf/lics/Keskin024}, we get the following corollary.
\begin{corollary}
        \regsep for bidirected VASS is \ackermanncomplete.
\end{corollary}
We remark that one can obtain a similar statement to Lemma~\ref{lem:bidirected} for other bidirected models, including bidirected \zvass. Moreover, the product VASS of two labeled bidirected VASSs is not necessarily a bidirected VASS. In fact, it is easy to show that the class of languages they represent is not closed under intersection. Observe that crucially in the proof of Lemma \ref{lem:bidirected} we used two different letters in each of the VASSs. Therefore, in the remaining part of the paper, we consider alphabetically bidirected systems. For alphabetically bidirected VASSs, their product is still such a VASS. This is formalised in \Cref{clm:product}, which is proven in Appendix~\ref{app:attempt}. Here for two VASSs $V_1 = (\Sigma_1,Q_1, \delta_1, q_{i_1}, q_{f_1})$ and $V_2 = (\Sigma_2,Q_2, \delta_2, q_{i_2}, q_{f_2})$, $d_1$ and $d_2$ dimensional respectively, we define their product $V_1 \times V_2 = (\Sigma, Q, \delta, q_i, q_f)$, which is a $d_1+d_2$ dimensional VASS, where $\Sigma = \Sigma_1 \cup \Sigma_2$, $Q = Q_1 \times Q_2$, $q_i = (q_{i_1}, q_{i_2})$, $q_f = (q_{f_1}, q_{f_2})$ and 
\[
\delta \;=\;
\Bigl\{
  \bigl( (p_1,p_2),\, a,\, (\myvec v_1, \myvec v_2),\, (q_1,q_2) \bigr)
  \;\Big|\;
  (p_i, a, \myvec v_i, q_i) \in \delta_i 
  \text{ for all } i \in \{1,2\}
\Bigr\}.
\]
\begin{claim}\label{clm:product}
    Let $V_1$ and $V_2$ be alphabetically bidirected VASSs over  $\Sigma$, then $V_1 \times V_2$ is also an alphabetically bidirected VASS over  $\Sigma$.
\end{claim}

	\section{Integer VASS}\label{sec:zvass}
We now study regular separability in alphabetically bidirected
\zvass. We will characterise its complexity via \emph{solvability of
integer linear Diophantine equation systems} (ILD), where given a matrix
$A\in\Z^{m\times n}$ and a vector $\vec{b}\in\Z^m$, we are asked whether
there is an $\vec{x}\in\Z^m$ with $A\vec{x}=\vec{b}$. ILD is solvable in
polynomial time~\cite[Thms.~1\&13]{chou1982algorithms}. By coILD, we denote the
complement of ILD, thus non-solvability of integer linear Diophantine equation
systems.
\begin{theorem}\label{thm:zvass}
 Regular separability for alphabetically bidirected \zvass is log-space
	interreducible with with coILD. In particular, the problem is in
	\ptime.
\end{theorem}
To show this, we will prove that regular separability for alphabetically
bidirected \zvass is log-space interreducible with emptiness in bidirected
\zvass, which is itself log-space interreducible with
coILD~\cite[Thm.~3.2(2)]{DBLP:conf/lics/GanardiMZ22}. The reduction from emptiness is
the usual one: Given a bidirected \zvass $V$ over $\Sigma$, we take the trivial
bidirected \zvass $V'$ with $L(V')=\Sigma^*$. Then $ L(V)=\emptyset$ if and
only if $L(V)\regsepop L(V')$. 

It thus remains to reduce regular separability to emptiness of bidirected
\zvass. To this end, we rely on the following lemma, which states that regular
separability is equivalent to intersection emptiness for \zvass.
\begin{restatable}{lemma}{zvassEquivalence}\label{lem:zvass_equivalence}
Alphabetically bidirected \zvass $V_1$ and $V_2$ are regularly separable if and only if $L(V_1) \cap L(V_2) = \emptyset$.
\end{restatable}
Note that this equivalence between disjointness and separability does not hold for general \zvass: The languages
$\{a^nb^m \mid n>m\}$, $\{a^nb^{m} \mid n<m\}$
are disjoint \zvass languages, but they are not separable by a regular language.

Before proving Lemma \ref{lem:zvass_equivalence}, we show how it implies Theorem~\ref{thm:zvass}.
\begin{proof}[Proof of \Cref{thm:zvass}]
	As argued, it remains to reduce regular separability to emptiness of bidirected \zvass. Let us fix \zvass $V_1$ and $V_2$. Consider the product \zvass $V_1 \times V_2$, which is
	alphabetically bidirected by Claim~\ref{clm:product}. Clearly
	$L(V_1 \times V_2) = L(V_1) \cap L(V_2)$. Thus, we have reduced the
	problem to emptiness of $L(V_1 \times V_2)$, which in turn log-space
	reduces to coILD~\cite[Thm.~3.2(2)]{DBLP:conf/lics/GanardiMZ22}.
\end{proof}

\subparagraph*{Overview of the proof of Lemma~\ref{lem:zvass_equivalence}}
The rest of this section is devoted to the proof of Lemma~\ref{lem:zvass_equivalence}. Firstly, we follow ideas from \cite{z_vass_separability_classical} (Lemmas \ref{lem:det_zvass}-\ref{lem:zvass_skeleton}), namely, we reduce regular separability to recognisable separability of semilinear sets in $\N^d$ (for some $d$). Moreover, in some sense, reductions preserve nonemptiness of languages and sets. Concretely speaking, instead of asking for the regular separability in two given \zvass, we are counting the cycles within runs of these \zvass. Finally, in order to finish the proof of Lemma~\ref{lem:zvass_equivalence}. We use Lemma~\ref{lem:zvass_sets} from \cite{new_zvass}, which characterises when two hyperlinear sets can be separated by a recognisable set. 

\subparagraph*{Reduction to deterministic systems}
We first follow the reduction from \cite{z_vass_separability_classical}. Firstly, we reduce in polynomial time the regular separability problem for nondeterministic alphabetically bidirected \zvass to the same problem in deterministic alphabetically bidirected \zvass. The fact that the reduction is performed in polynomial time, while determinization typically requires at least an exponential blowup, follows from the fact that in our "determinization" we use a homomorphism. Concretely speaking, from two given alphabetically bidirected \zvass $V_1$ and $V_2$ over $\Sigma$ we construct two deterministic alphabetically bidirected VASS $W_1$ and $W_2$ $\Gamma$ such that $L(W_i) = h^{-1}(L(V_i))$ for some homomorphism $h: \Sigma^* \rightarrow \Gamma^*$ and $L(V_1) \regsepop L(V_2)$ if, and only if, $L(W_1) \regsepop L(W_2)$ holds. Moreover, $L(V_1) \cap L(V_2) = \emptyset$ if, and only if, $L(W_1) \cap L(W_2) = \emptyset$. This idea is analogous to \cite[Lemma~7]{z_vass_separability_classical}, but we need to make sure that bidirectedness is preserved. A proof can be found in \cref{app:det_zvass}.
\begin{restatable}{lemma}{detZVASS}\label{lem:det_zvass}
    Given alphabetically bidirected \zvass $V_1, V_2$, one can construct
	deterministic alphabetically bidirected \zvass $W_1, W_2$ such that
	\begin{enumerate}
		\item $L(W_1)\regsepop L(W_2)$ if and only if $L(V_1)\regsepop L(V_2)$ and
		\item $L(W_1) \cap L(W_2) = \emptyset$ if and only if $L(V_1) \cap L(V_2) = \emptyset$.
	\end{enumerate} 
\end{restatable}
This means, it suffices to prove \cref{lem:zvass_equivalence} for deterministic
alphabetically bidirected \zvass.

\subparagraph*{Reduction to a single integer VASS} 
As in \cite{z_vass_separability_classical}, we now simplify the situation further. Concretely speaking, we reduce it to regular separability of two languages accepted by the same deterministic alphabetically bidirected \zvass, but with different sets of
counters, that is given $d_1$-\zvass $V_1$ and $d_2$-\zvass $V_2$ we construct (using product construction) one $(d_1+d_2)$-\zvass $V$ and two index sets $I_1, I_2 \subseteq [1, d_1+d_2]$ such that $L(V_i) = L(V, I_i)$ where $L(V, I_i)$ is the language of the VASS $V$ restricted only to the counters from the subset $I_i$. Thus, clearly $L(V_1) \cap L(V_2) = L(V, I_1) \cap L(V, I_2)$. Note that, as we use product construction, the fact that we work with alphabetically bidirected \zvass is crucial as we can use Claim~\ref{clm:product} and say that $V$ is also a bidirected \zvass over a bidirected alphabet. The following lemma is analogous to \cite[Proposition~1]{z_vass_separability_classical}, but in the bidirected setting and is proven in \cref{app:one_zvass}.
\begin{lemma}\label{lem:one_zvass}
	Given deterministic alphabetically bidirected \zvass $V_1$ and $V_2$,
	one can construct a single deterministic, alphabetically bidirected
	$d$-\zvass $V$ with subsets $I_1,I_2\subseteq[1,d]$ such that
	\begin{enumerate}
		\item $L(V,I_1)\regsepop L(V,I_2)$ if and only if $L(V_1)\regsepop
	L(V_2)$ and 
\item $L(V,I_1)\cap L(V,I_2)=\emptyset$ if and only if
	$L(V_1)\cap L(V_2)=\emptyset$.
	\end{enumerate}
\end{lemma}
This means, it suffices to prove \cref{lem:zvass_equivalence} for a language
pair of the form $L(V,I_1),L(V,I_2)$ as in \cref{lem:one_zvass}.

\subparagraph*{Skeletons}
Further, we simplify the regular separability problem by
considering only runs in $V$ that are in some sense similar. Concretely speaking, we consider some base paths in $V$, which we call \emph{skeletons}. We say that two runs in $V$ are similar if they follow the same base
path and only differ in the order and repetition of some cycles. Concretely, we define a function $\skel: T^* \rightarrow T^*$  satisfying that for a path $\pi \in T^*$ we have $\skel(\pi) = \rho$ such that (i) $\rho$ is a sub-path of $\pi$ (ii) 
the set of states visited by $\pi$ is the same as the set of states visited by $\rho$ and (iii) the last state of $\pi$ is the same as the last state of $\rho$. We call $\rho$ the skeleton of $\pi$. The definition of the function $\skel: T^* \rightarrow T^*$ is standard and closely follows the one, which can be found in \cite{z_vass_separability_classical, new_zvass}. For completeness, we provide it in \cref{app:skeletons}.   
By \cite[Lemma~10]{z_vass_separability_classical} we get, that for any path $\pi$ it holds that  $|\skel(\pi)| \leq |T|^2$. For a skeleton $\rho$, we say that a $\rho$-cycle is a simple cycle that only visits states occurring in $\rho$; a $\rho$-run is
a run $\pi$ such that $\skel(\pi) = \rho$. Observe that therefore every $\rho$-run can be obtained from $\rho$ by inserting $\rho$-cycles. Finally, we
define $L(V, I, \rho)$ as the set of all words from $L(V, I)$ accepted by $\rho$-runs. 
\begin{lemma}[{\cite[Lemma~11]{z_vass_separability_classical}}]\label{lem:zvass_skeleton}
    We have $L(V, I_1) \regsepop L(V, I_2)$ if, and only if, $L(V, I_1, \rho) \regsepop
L(V, I_2, \rho)$ holds for every skeleton $\rho$.
\end{lemma}
Since clearly $L(V,I_1)\cap L(V,I_2)=\emptyset$ if and only if
$L(V,I_1,\rho)\cap L(V,I_2,\rho)=\emptyset$ for every skeleton, this means it
suffices to prove \cref{lem:zvass_equivalence} for language pairs of the form
$L(V,I_1,\rho),L(V,I_2,\rho)$.

\subparagraph*{Counting cycles}
Now, we recall a characterisation of regular separability from \cite{z_vass_separability_classical}. It says that regular separability of the languages $L(V, I_1, \rho)$ and $L(V, I_2, \rho)$ is equivalent to
recognisable separability of vectors that count $\rho$-simple cycles. This is possible since each $\rho$-cycle can be decomposed into simple $\rho$-cycles. In the following, we fix the set $S$ of all simple $\rho$-cycles in $V$, which is a finite set. For a multiset of simple cycles $\vec{u} \in \N^{|S|}$  and $I \subseteq [1,d]$ we define an effect of $\vec{u}$ on the components in $I$ as $\eff_I(\vec{u}) = \sum_{c \in S}\vec{u}(c) \cdot \eff_I(c)$. Similarly for a multiset of transitions $\vec{v} \in \N^{|T|}$ and $I \subseteq [1,d]$ we define an effect of $\vec{v}$ on the components in $I$ as $\eff_I(\vec{v}) = \sum_{t \in T} \vec{v}(t) \cdot \eff_I(t)$. 
Finally, we define
$$M(V,I,\rho) = \setof{\vec{u} \in \N^{|S|}}{\eff_I(\rho) + \eff_I(\vec{u}) = \vec 0}$$

Hence, $M(V, I, \rho)$ is the set of multisets of simple $\rho$-cycles such that inserting them into $\rho$ would lead to an accepting run with acceptance condition $I \subseteq [1, d]$. The following equivalence between regular separability of the languages $L(V, I_i, \rho)$ and
recognisable separability of the sets $M(I_i)$ was shown in \cite{z_vass_separability_classical}.
\begin{lemma}[{\cite[Lemma~12]{z_vass_separability_classical}}]\label{lem:zvass_sets}
    $L(V, I_1, \rho) \regsepop L(V, I_2, \rho)$ if and only if $M(V, I_1, \rho) \regsepop M(V, I_2, \rho)$.
\end{lemma}
This means, for \cref{lem:zvass_equivalence}, it suffices to show:
\begin{equation} M(V,I_1,\rho)\cap M(V,I_2,\rho)=\emptyset~~\Longrightarrow~~M(V,I_1,\rho)\regsepop M(V,I_2,\rho). \label{zvass-remains} \end{equation}
	Indeed, by the above, we need to show that $L(V,I_1,\rho)\cap L(V,I_2,\rho)=\emptyset$ implies $L(V,I_1,\rho)\regsepop L(V,I_2,\rho)$. But if $L(V,I_1,\rho)\cap L(V,I_2,\rho)=\emptyset$, then certainly $M(V,I_1,\rho)\cap M(V,I_2,\rho)=\emptyset$, which by \eqref{zvass-remains} and \cref{lem:zvass_sets} would yield $L(V,I_1,\rho)\regsepop L(V,I_2,\rho)$.  Thus, we shall prove \eqref{zvass-remains}.

\subparagraph*{Hyperlinear sets} Now we deviate from the ideas from
$\cite{z_vass_separability_classical}$ and use ideas from \cite{new_zvass}.
Firstly, the set $M(V, I, \rho)\subseteq\N^{|S|}$ is hyperlinear: 
\begin{claim}\label{clm:hyperlinear}
	Let $I \subseteq [1, d]$. Then $M(V, I, \rho)$ is hyperlinear, i.e., $M(V, I, \rho) = A + X^*$ for two finite sets $A, X \subseteq \N^{|S|}$
\end{claim}
This is because, we can take as $A$ the minimal solutions $\vec{u}\in\N^{|S|}$ to $\eff_I(\rho)+\eff_I(\vec{u})=\vec{0}$, of which there are finitely many. Moreover, the set
\[ H_I = \{\vec{u}\in\N^{|S|} \mid \eff_I(\vec{u})=\vec{0}\} \]
satisfies $M(V,I,\rho)=A+H_I$, and is generated as a monoid by its minimal elements, meaning $H_I=X^*$ for some finite $X\subseteq\N^{|S|}$ (see \cite[Observation 6.4]{new_zvass} and \cite{semilinear}). 

\subparagraph*{Separating hyperlinear sets}
Furthermore, we will rely on a characterisation of recognisable separability of hyperlinear sets from \cite{new_zvass}. To state it, we need some terminology. For a vector $\vec{u}\in\N^k$, we write $\supp(\vec{u})=\{j\in [1,k]\mid \vec{u}(j)\ge 1\}$. A coordinate $j \in [1, k]$ is \emph{twin-unbounded} for \( K = A + U^* \subseteq \mathbb{N}^k \) and \( L = B + V^* \subseteq \mathbb{N}^k \) if there exist $\vec{p} \in U^*$ and $\vec{q} \in V^*$ such that $j \in \supp(\vec{p}) = \supp(\vec{q})$. Moreover, for $J \subseteq [1,k]$, we write $U_J = \{\vec{p} \in U \mid \supp(\vec{p}) \subseteq J\}$ and $V_J = \{\vec{q} \in V \mid \supp(\vec{q}) \subseteq J\}$.

\begin{lemma}[{\cite[Theorem~4.6]{new_zvass}}]\label{lem:sep_characterization}
Let \( K = A + X^* \subseteq \mathbb{N}^k \) and \( L = B + Y^* \subseteq \mathbb{N}^k \) be hyperlinear sets. 
Then \( K \) and \( L \) are not separable by a recognisable set if and only if the intersection
\[
(A + X^* + Y_J^*) \cap (B + Y^* + X_J^*)
\]
is non-empty, where \( J \subseteq [1, k] \) is the set of coordinates that are twin-unbounded for \( K, L \)
\end{lemma}

\subparagraph*{Exploiting bidirectedness} Until now, our reductions worked
essentially like the reductions in either \cite{z_vass_separability_classical}
or \cite{new_zvass}, respectively. This means, they preserve bidirectedness,
but we have not exploited bidirectness yet. Our next lemma finally exploits bidirectedness.
\begin{proposition}\label{lem:homogeneous-vector}
	If $V$ is bidirected, then there is a $\vec{w}\in H_{I_1}\cap H_{I_2}$
	with $\supp(\vec{w})=[1,d]$.
\end{proposition}
\begin{proof}[Proof sketch]
	(A full proof is in \cref{app:zvass_equivalence}.) Since every cycle $\pi\in
	S$ has a backward cycle $\bar{\pi}\in S$, the vector $\vec{w}_{\pi}$,
	which contains $\pi$ and $\bar{\pi}$ (but no other cycle) will have
	$\eff_{[1,d]}(\vec{w}_{\pi})=\vec{0}$. With $\vec{w}=\sum_{\pi\in S}
	\vec{w}_\pi$, we get a vector as desired.
\end{proof}

With \cref{lem:sep_characterization,lem:homogeneous-vector} in hand, we can now
prove \eqref{zvass-remains}. 

\begin{proof}[Proof of \eqref{zvass-remains}]
	We write $M(V,I_1,\rho)=A+X^*$ and $M(V,I_2,\rho)=B+Y^*$ for finite
	sets $A,B,U,V\subseteq\N^{|S|}$, and let $J\subseteq[1,|S|]$ be the set
	of twin-unbounded coordinates. Then we have $X^*=H_{I_1}$ and $Y^*=H_{I_2}$. We
	prove the contraposition of \eqref{zvass-remains}, so suppose
	$M(V,I_1,\rho)\regsepop M(V,I_2,\rho)$ does not hold. By
	\cref{lem:sep_characterization} we have 
\[ \vec{a}+\vec{x}-\vec{x}_J=\vec{b}+\vec{y}-\vec{y}_J \]
for some $\vec{a}\in A$, $\vec{x}\in U^*$, $\vec{x}_J\in U_J^*$, $\vec{b}\in
	B$, $\vec{y}\in V^*$, and $\vec{y}_J\in V_J^*$. With the vector
	$\vec{w}\in H_{I_1}\cap H_{I_2}$ from \cref{lem:homogeneous-vector} and some
	sufficiently high $k\in\N$, we have $k\cdot\vec{w}-\vec{x}_J\ge\vec{0}$
	and $k\cdot\vec{w}-\vec{y}_J\ge\vec{0}$, since $\vec{w}$ is
	positive in every coordinate. We claim that the vector 
\begin{equation} \vec{a}+\vec{x}+k\cdot\vec{w}-\vec{x}_J=\vec{b}+\vec{y}+k\cdot\vec{w}-\vec{y}_J \label{vector-intersection}\end{equation}
belongs to $M(V,I_1,\rho)\cap M(V,I_2,\rho)$: Indeed, since
	$\vec{x}_J,\vec{w}\in H_{I_1}$ and also $k\cdot\vec{w}-\vec{x}_J\ge\vec{0}$, we know that $k\cdot\vec{w}-\vec{x}_J\in H_{I_1}=X^*$. Therefore,
	\eqref{vector-intersection} belongs to $M(V,I_1,\rho)$. A symmetric
	argument yields membership in $M(V,I_2,\rho)$.
\end{proof}
A detailed proof with all steps in order can be found in \cref{app:zvass_equivalence}.

	\section{VASS}\label{sec:vass}
In this section, we prove the following:
\begin{theorem}\label{thm:vass}
Regular separability for alphabetically bidirected VASS is \expspacecomplete.
\end{theorem}
We proceed similarly as in Section~\ref{sec:zvass}, that means we show Lemma~\ref{lem:vass_equivalence}, which states that regular separability is equivalent to intersection emptiness for VASS.
\begin{lemma}\label{lem:vass_equivalence}
Alphabetically bidirected VASS $V_1$ and $V_2$ are regularly separable if and only if $L(V_1)\cap L(V_2) = \emptyset$.
\end{lemma}
Before proving Lemma~\ref{lem:vass_equivalence} we show how it implies Theorem~\ref{thm:vass},
\begin{proof}[Proof of Theorem~\ref{thm:vass}]
For \expspace-hardness, we can use the usual reduction from reachability, since
	reachability in bidirected VASS is \expspacecomplete~\cite[Main
	Theorem]{mayr1982complexity}. Suppose we are given a bidirected VASS
	$V$ and want to decide whether $q_i(\vec{0})$ can reach $q_f(\vec{0})$.
	We construct an alphabetically bidirected VASS $V_1$ whose input words
	are transition sequences of $V$ that lead from $q_i(\vec{0})$ to
	$q_f(\vec{0})$. With an alphabetically bidirected $V_2$ that accepts
	all words, we then have $L(V_1)\regsepop L(V_2)$ if and only if
	$q_f(\vec{0})$ is not reachable from $q_i(\vec{0})$. A detailed proof is in \cref{app:hardness}.

	For the $\expspace$ upper bound, fix alphabetically bidirected
	VASSs $V_1$ and $V_2$. Because of Lemma~\ref{lem:vass_equivalence} we
	need only to check whether $L(V_1) \cap L(V_2) = \emptyset$. In
	polynomial time, we construct product VASS $V_1 \times V_2$, which,
	because of Claim~\ref{clm:product}, is an alphabetically bidirected
	VASS. Clearly $L(V_1 \times V_2) = L(V_1) \cap L(V_2)$. Hence, we need
	to check the non-emptiness of $L(V_1 \times V_2)$, which reduces to a
	reachability query, which is decidable in $\expspace$.
\end{proof}

We now prove Lemma~\ref{lem:vass_equivalence}. The
non-obvious direction is the implication that disjointness implies regular
separability. Therefore, fix $d_1$-VASS $V_1$ and $d_2$-VASS $V_2$ such that
\[ L(V_1)\cap L(V_2)=\emptyset. \]

\subparagraph{Separability of reversal languages}
In the following, we will sometimes argue about the reversal VASS $\rev{V}$ of
$V$. 
Here, for a language $L\subseteq\Sigma^*$, we denote by $\rev(L)$ the
\emph{(bidirected) reversal language of $L$}, meaning $\rev(L)=\{w^{\rev} \mid w\in L\}$,
where $w^{\rev}$ is the reversal of the word $w$, where also all letters are
turned into their opposite (i.e.\ $a$ is replaced with $\bar{a}$ and
vice-versa). Then we have $L(\rev{V})=\rev(L(V))$.
We will rely on the following simple observation (see \cref{app:vass_1} for a proof):
\begin{claim}\label{clm:reverse}
    For any languages $L_1,L_2$, we have $L_1\regsepop L_2$ if and only if
	$\rev(L_1)\regsepop \rev(L_2)$.
\end{claim}

\subparagraph{Case I: Simultaneous unboundedness}
We first consider a special case when the pairs $(V_1, V_2)$ and $(\rev(V_1), \rev(V_2))$ are simultaneously unbounded. Formally, a pair of VASSs $(V_1, V_2)$ is \emph{simultaneously unbounded} if for the initial state $q_i$ of $V_1 \times V_2$, all states $q$ of $V_1 \times V_2$ and every $C \in \N$ there exists $\myvec u \in \reach(V,q_i(\vec 0), q)$ such that, for every $j \in [1, d_1 + d_2]$, we have $\myvec u_j > C$. In other words, for every $C \in \N$, all counters in $V_1 \times V_2$ can go above $C$ at the same time. 

\subparagraph{Coverability overapproximation}
Our arguments crucially depend on considering overapproximations of $L(V_1)$ and $L(V_2)$.
For a VASS $V$, let $L_{\cover}(V)$ be its \emph{coverability language}, meaning the set of all words read by runs that end up in a final state, \emph{with any counter values}.
Observe, that $L(V_i) \subseteq L_{\cover}(V_i)$ and $L(\rev(V_i)) \subseteq L_{\cover}(\rev(V_i))$. Thus, regular separability of $L_{\cover}(V_1)$ and $L_{\cover}(V_2)$ would imply regular separability of $L(V_1)$ and $L(V_2)$. Similarly, because of Claim~\ref{clm:reverse}, regular separability of $L_{\cover}(\rev(V_1))$ and $L_{\cover}(\rev(V_2))$ would imply regular separability of $L(V_1)$ and $L(V_2)$. If one of this happen, we are done showing regular separability of $L(V_1)$ and $L(V_2)$. Therefore, assume that neither of these two situations happens. Then we use the following lemma from \cite[Theorem~3.1]{regular}.
\begin{lemma}[{\cite[Theorem~3.1]{regular}}]\label{lem:coverability}
    For two VASSs $V_1$ and $V_2$ we have $L_{\cover}(V_1) \regsepop L_{\cover}(V_2)$ if and only if $L_{\cover}(V_1) \cap L_{\cover}(V_2) = \emptyset$.\footnote{In fact, this result was shown for a wider class of Well-structured transition systems (WSTS).}
\end{lemma}
Therefore due to Lemma~\ref{lem:coverability} we have $L_{\cover}(V_1) \cap L_{\cover}(V_2) \neq \emptyset$ and $L_{\cover}(\rev(V_1)) \cap L_{\cover}(\rev(V_2)) \neq \emptyset$. 

\subparagraph*{Constructing runs}
We now use the non-emptiness of these intersections, and simultaneous unboundedness, to construct certain runs of $V_1$ and $V_2$.
Intuitively speaking, the property of the words, which we look for, is that they can be accepted by a run having a strictly positive effect on all the counters. Let $q_i$ and $p_i$ be the initial and final state of $V_i$ respectively. Then we can observe the following.
\begin{claim}\label{clm:covering_word}
    There exist $u_1 \in L_{\cover}(V_1) \cap L_{\cover}(V_2)$ and $u_2 \in  L_{\cover}(\rev(V_1)) \cap L_{\cover}(\rev(V_2))$  such that 
    for each $i \in [1,2]$ there exist runs:
    \begin{itemize}
        \item $\rho_{u_1}^i$ in $V_i$ over $u_1$, which when fired from configuration $q_i(\vec{0})$ covers configuration $p_i(\vec{1})$ and with $\eff(\rho_{u_1}^i) \geq \vec{1}$
        \item $\rho_{u_2}^i$ in $V_i$ over $u_2$, which when fired from configuration $p_i(\vec{0})$ covers configuration $q_i(\vec{1})$ and with $\eff(\rho_{u_2}^i) \geq \vec{1}$.
    \end{itemize}
\end{claim}
Let us give some intuition for the first statement. Because of simultaneous
unboundedness, there must be a state $q$ and two reachable configurations
$q(\vec{v})$ and $q(\vec{v}')$ such that $\vec{v'}$ is larger than $\vec{v}$ on
all coordinates. By first going to $q(\vec{v}')$ and then (exploiting
bidirectedness) taking the path to $q(\vec{v})$ backwards, we can come back to
the initial state, but with all counters positive. After this, we apply the
run given by $L_{\cover}(V_1)\cap L_{\cover}(V_2)\ne\emptyset$. A detailed
proof of \Cref{clm:covering_word} can be found in Appendix~\ref{app:vass_2}. 

\subparagraph*{Proving regular separability} Now, we show that $L(V_1)$ and
$L(V_2)$ are regularly separable. To this end, we will show that $L_{\Z}(V_1)$
and $L_{\Z}(V_2)$ are disjoint. This suffices, because $L(V_i) \subseteq
L_{\Z}(V_i)$ and by Lemma~\ref{lem:zvass_equivalence}, this means $L(V_1)$ and
$L(V_2)$ are regularly separable.

Assume, towards contradiction that $L_{\Z}(V_1) \cap L_{\Z}(V_2) \neq \emptyset$. Therefore, there exist  $\Z$-run $\rho_1$ and $\rho_2$ in $V_1$ and $V_2$, respectively, which are accepting runs (in the $\Z$-semantics) and read the same word $w$. Our goal is to come up with a contradiction that  $L(V_1) \cap L(V_2) = \emptyset$. Let $M$ be the maximal change of a counter value by a single transition in $V_1$ and $V_2$ and let $k = \max(|\rho_1|, |\rho_2|)$. For $i \in [1,2]$, let us define run 
$$\pi_i = (\rho_{u_1}^i\rho_{u_2}^i)^{Mk}\rho_i[(\overline{\rho_{u_1}^i})(\overline{\rho_{u_2}^i})]^{Mk}$$
Observe that $\pi_i$ is a $\Z$-run in $V_i$. Moreover we have, that $\pi_1$ and $\pi_2$ read the same word $u$ and $\eff(\pi_i) = \eff(\rho_i) = \vec{0}$. Now we show that they are runs starting from $q_i(\vec{0})$. Clearly $(\rho_{u_1}^i\rho_{u_2}^i)^{Mk}$ starting from $q_i(\vec{0})$ is a run and observe, that $\eff((\rho_{u_1}^i\rho_{u_2}^i)^{Mk}) \geq Mk \geq M|\rho_i|$ and hence $(\rho_{u_1}^i\rho_{u_2}^i)^{Mk}\rho_i$ is a run starting from $q_i(\vec{0})$. As $(\rho_{u_2}^i\rho_{u_1}^i)^{MK}$ is a run starting from $p_i(\vec 0)$ to $p_i(\myvec v)$, for $\myvec v \geq \vec{0}$ we get that $[(\overline{\rho_{u_1}^i})(\overline{\rho_{u_2}^i})]^{Mk}$ is a run from $p_i(\myvec v)$ to $p_i(\vec{0})$. Observe that $\eff((\rho_{u_1}^i\rho_{u_2}^i)^{Mk}\rho_i) = \myvec v$ and hence
$\pi_i$ is an accepting run in $V_i$ and therefore we get contradiction with $L(V_1) \cap L(V_2) = \emptyset$, which results in $L_{\Z}(V_1) \cap L_{\Z}(V_2) = \emptyset$, which, because of Lemma~\ref{lem:zvass_equivalence}, implies that $L(V_1)$ and $L(V_2)$ are regularly separable. The above considerations can be summarised in the following claim.
\begin{claim}\label{clm:vass_regsep_easy}
    If $L(V_1) \cap L(V_2) = \emptyset$ and $(V_1, V_2)$ and $(\rev(V_1), \rev(V_2))$ are simultaneously unbounded then $L(V_1) \regsepop L(V_2)$. 
\end{claim}

\subparagraph*{The general case}
We now show the general case of \Cref{lem:vass_equivalence}, i.e.\ where we do not assume simultaneous unboundedness. The full proof can be found in Appendix \ref{app:vass_3}. We proceed by induction on $d_1+d_2$. If $d_1 + d_2 = 0$ then the property of Lemma~\ref{lem:vass_equivalence} is clearly true because  $L(V_1)$ and $L(V_2)$ are regular languages. Now assume that $d_1 + d_2 > 0$. If $(V_1, V_2)$ and $(\rev(V_1), \rev(V_2))$ are simultaneously unbounded then, by \Cref{clm:vass_regsep_easy}, we get $L(V_1) \regsepop L(V_2)$, as required. Otherwise, because of \Cref{clm:reverse}, we can w.l.o.g. assume that $(V_1, V_2)$ is not simultaneously unbounded. In this case we leverage the following lemma from \cite{DBLP:conf/lics/GanardiMZ22}.

\begin{lemma}[{\cite[Lemma~6.2]{DBLP:conf/lics/GanardiMZ22}}]\label{lem:partition}
    For every bidirected VASS $V$ with source configuration $s$ there exists a number $B \in \N$ and we can partition the set of counters into two disjoint subsets of counters $P_1, P_2$ such that for all states $q$ we have:
    \begin{itemize}
        \item for every $\myvec v \in \reach(V,s,q)$ and every $i \in P_1$ we have $\myvec v_i \leq B$;
        \item for every $C \in \N$ there is a $\myvec u \in \reach(V, s, q)$ with $\myvec u_i > C$ for every $i \in P_2$.
    \end{itemize}
\end{lemma}

Let us consider VASS $V_1 \times V_2$, which is an alphabetically bidirected
VASS due to Claim~\ref{clm:product} and partition its counters into two sets
$P_1$ and $P_2$ as in Lemma~\ref{lem:partition}. Since $(V_1, V_2)$ is not
simultaneously unbounded we get that $P_1 \neq \emptyset$. \Wlog assume that
there exists counter $i$ coming from $V_1$ such that $i \in P_1$, meaning
counter $i$ is bounded $B$ in $V_1\times V_2$. 

\subparagraph{Splitting along counter bound}
However, note that while counter $i$ is bounded in $V_1\times V_2$, it may not
be bounded in $V_1$. Therefore, from $V_1$, we now construct two VASS:
$V_{1,\le B}$, which simulates all runs of $V_1$ where counter $i$ remains at
most $B$; and $V_{1,>B}$, which simulates all runs of $V_1$ where counter $i$
exceeds $B$ at some point. This means, $V_{1,>B}$ goes into a special accepting
state from which it can read any letter. Crucially, tracking the value of
counter $i$ can be done in a control state, meaning $V_{1,\le B}$ and
$V_{1,>B}$ have one fewer counter (and we will be able to apply induction).

Moreover, we observe that $V_{1,\le B}$ is alphabetically bidirected, whereas
$V_{1,>B}$ may not be. However, we can make $V_{1,>B}$ alphabetically
bidirected by explicitly adding the reverse transitions, without changing its
language. 

We can now observe that both $L(V_{1,\le B})$ and $L(V_{1,>B})$ are disjoint
from $L(V_2)$, for different reasons: $L(V_{1,\le B})$ is disjoint from
$L(V_2)$, because any intersection would also exist in $L(V_1)$ and $L(V_2)$.
Moreover, $L(V_{1,>B})$ is disjoint from $L(V_2)$ because of counter $i$ being
bounded by $B$ in $V_1\times V_2$. For both languages, induction yields that
they are regularly separable from $L(V_2)$. Since $L(V_1) \subseteq L(V_{1,\le
B}) \cup L(V_{1,>B})$, this means $L(V_1)$ and $L(V_2)$ are also regularly
separable, and \Cref{lem:vass_equivalence} is established.

	\section{OCA}\label{sec:undecidability}
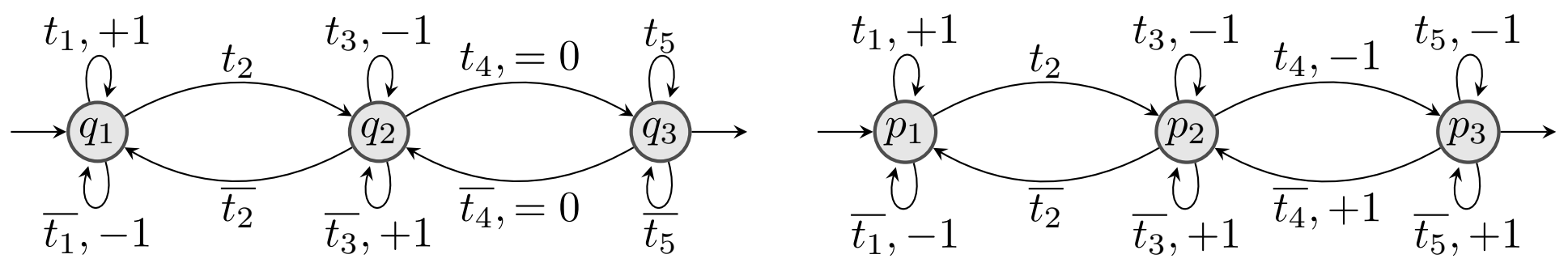
\begin{figure}
    \centering
    \begin{tikzpicture}[node distance=2.3cm, gadget]
        \node[state,initial,initial text=] (A1) {$q_1$};
        \node[state, right of=A1] (A2) {$q_2$};
        \node[state, right of=A2,accepting by arrow] (A3) {$q_3$};

        \draw[->] (A1) edge[loop above] node{$t_1, +1$} (A1);
        \draw[->] (A2) edge[loop above] node{$t_3, -1$} (A2);
        \draw[->] (A3) edge[loop above] node{$t_5$} (A3);
        \draw[->] (A1) edge[loop below] node{$\overline{t_1}, -1$} (A1);
        \draw[->] (A2) edge[loop below] node{$\overline{t_3}, +1$} (A2);
        \draw[->] (A3) edge[loop below] node{$\overline{t_5}$} (A3);

        \draw[->] (A1) edge[bend left] node[above]{$t_2$} (A2);
        \draw[->] (A2) edge[bend left] node[below]{$\overline{t_2}$} (A1);
        \draw[->] (A2) edge[bend left] node[above]{$t_4, =0$} (A3);
        \draw[->] (A3) edge[bend left] node[below]{$\overline{t_4}, =0$} (A2);

        \node[state, right=2cm of A3,initial,initial text=] (B1) {$p_1$};
        \node[state, right of=B1] (B2) {$p_2$};
        \node[state, right of=B2,accepting by arrow] (B3) {$p_3$};

        \draw[->] (B1) edge[loop above] node{$t_1, +1$} (B1);
        \draw[->] (B2) edge[loop above] node{$t_3, -1$} (B2);
        \draw[->] (B3) edge[loop above] node{$t_5, -1$} (B3);
        \draw[->] (B1) edge[loop below] node{$\overline{t_1}, -1$} (B1);
        \draw[->] (B2) edge[loop below] node{$\overline{t_3}, +1$} (B2);
        \draw[->] (B3) edge[loop below] node{$\overline{t_5}, +1$} (B3);
        \draw[->] (B1) edge[bend left] node[above]{$t_2$} (B2);
        \draw[->] (B2) edge[bend left] node[below]{$\overline{t_2}$} (B1);
        \draw[->] (B2) edge[bend left] node[above]{$t_4, -1$} (B3);
        \draw[->] (B3) edge[bend left] node[below]{$\overline{t_4}, +1$} (B2);
        \end{tikzpicture}
    \caption{Example of OCA $V_1$ (on the left) and OCA $V_2$ (on the right) satisfying the conditions of Lemma~\ref{lem:conterexample}. For $V_1$, we set its initial state to $q_1$ and its final state to $q_3$. For $V_2$, we set its initial state to $p_1$ and its accepting state to $p_3$. Annotations on transitions represent the letter of a transition and its effect. $+1, -1, =0$ mean, respectively, incrementing, decrementing, and zero-testing the counter.} 
    \label{fig:conterexample}
\end{figure}

In this section, we consider alphabetically bidirected OCA. In light of the results for bidirected \zvass and VASS, one might suspect that, for bidirected systems, intersection non-emptiness is always equivalent to regular separability. However, this is not the case:

\begin{lemma}\label{lem:conterexample}
There exist alphabetically bidirected OCA $V_1$ and $V_2$ such that $L(V_1)$ and $L(V_2)$ are not regularly separable and $L(V_1) \cap L(V_2) = \emptyset$.
\end{lemma}
The OCA are shown in \cref{fig:conterexample}. Intuitively, their languages
are not regularly separable, because the first occurrence of $t_4$ requires the
counter, which is visible in the balance between positive (unbarred) and
negative (barred) letters, to be zero in $V_1$, but non-zero in $V_2$. This
cannot be distinguished by a regular language. A detailed proof can be found in
Appendix~\ref{app:undecidability}.

In fact, we will show the following:
\begin{theorem}\label{thm:undecidability}
Regular separability is undecidable for alphabetically bidirected OCA.
\end{theorem}
This strengthens undecidability of regular separability of general OCA, as shown by Czerwiński and Lasota in \cite{ocn_separability}. We adapt their proof to the bidirected setting. We show, roughly speaking, that every decidable problem can be reduced in exponential time to regular separability of languages of alphabetically bidirected OCA, which implies undecidability. This technique is inspired by the undecidability proof presented by Hunt \cite{hunt}.
The rest of this section is devoted to the proof of Theorem~\ref{thm:undecidability}.

\subparagraph*{Overall strategy}
Similar to the proof of Czerwiński and Lasota in \cite{ocn_separability}, our goal is to show a time-bounded reduction from any decidable problem $D \subseteq \{0,1\}^*$ to regular separability for alphabetically bidirected OCA. However, instead of reducing in polynomial time, we reduce in exponential time. This implies undecidability of the latter problem, since otherwise, decidability of regular separability for alphabetically bidirected OCA in time $f(n)$, where $n$ is the size of the input, would imply decidability of any decidable problem in time $f(2^{p(n)})$, for some polynomial $p$, which would violate the time hierarchy theorem \cite[Theorem~9.10]{time_hierarchy}. Hence, in order to prove Theorem~\ref{thm:undecidability}, it is enough to prove the following claim.
\begin{claim}\label{oca-reduction}
	Every decidable problem $D\subseteq\{0,1\}^*$ reduces in exponential time to regular separability of languages for alphabetically bidirected OCA.
\end{claim}

\subparagraph{Two-counter machines} 
A \emph{deterministic \twocm} consists of a finite set of states $Q$ with three distinguished states: initial state $q_i$, accepting state $q_a$ and rejecting state $q_r$; two counters $c_1$ and $c_2$; and a set of transitions, which can increment, decrement, or zero-test a counter. Crucially, counter values cannot drop below zero. Furthermore, the accepting and rejecting states $q_a$ and $q_r$ have no outgoing transitions. Finally, for every state $q \in Q$ and every pair of counter values $n_1,n_2$, at most one transition is enabled. A configuration $q(n_1,n_2)$ of $M$ consists of a state $q$ and values of counters $n_1, n_2$. For a vector $\myvec v \in \N^2$ we write $q(\myvec v)$ for configuration $q(\myvec v_1,\myvec v_2)$. For two configurations $p(\myvec v)$ and $q(\myvec u)$, we write \trans{p(\myvec v)}{q(\myvec u)} if there exists a sequence of transitions leading from $p(\myvec v)$ to $q(\myvec u)$, and $p(\myvec v) \xrightarrow{} q(\myvec u)$ if there exists a single transition leading from $p(\myvec v)$ to $q(\myvec u)$.

We say that a number $k \in \N$ is \emph{accepted} by \twocm $M$ if \trans{q_i(k,0)}{q_a(0,0)} and \emph{rejected} if \trans{q_i(k,0)}{q_r(0,0)}. We call $M$ \emph{total} if every $k \in \N$ is either accepted or rejected by $M$. The language of $M$, denoted by $L(M)$, is the set of numbers accepted by $M$.

\subparagraph*{Accepting languages} Let us encode every word $w \in \set{0,1}^*$ using binary representation as a natural number $n(w)$. It is well-known that then for every decidable language $D\subseteq\{0,1\}^*$,  there exists a total deterministic
\twocm $N$ such that $w \in D$ if, and only if $2^{n(w)} \in L(N)$. Here, the exponentiation arises from the fact that firstly a Turing Machine is simulated by a $3$-counter machine, which is later simulated by \twocm, which stores the value of three counters $c_1, c_2, c_3$ as the value $2^{c_1}3^{c_2}5^{c_3}$. 

\subparagraph*{Reversible Two-counter machines} 
In our proof, it will be crucial to work with \emph{reversible} two-counter machines, meaning ones where
for every configuration $q(\myvec v)$ there is at most one configuration $p(\myvec u)$ such that $p(\myvec u) \xrightarrow{} q(\myvec v)$. By a result of Morita~\cite[Theorem 4.1]{reversible_automata},
every total deterministic \twocm can be simulated by a total deterministic reversible \twocm. This means, we may assume that our machine $N$ is total deterministic and reversible.
Moreover, by adding auxiliary states and transitions, we can assume that $N$ has only a single outgoing transition from the initial state.

\subparagraph*{Adding backward edges} Let us now prove \cref{oca-reduction}. We
take a decidable set $D\subseteq\{0,1\}^*$ and construct $N$ as above. Let $T$
be the set of transitions in $N$. Recall that in $N$, there is only one
outgoing transition from the initial state of $N$. We construct a \twocm $M$
from $N$ by adding reverse edges, namely, for each transition incrementing,
decrementing, or zero-testing a counter, we add a reverse transition
decrementing, incrementing, or zero-testing the counter respectively.

We observe, that $L(M) = L(N)$. For that, we use an argument similar to the one
presented already in Post’s undecidability proof of the word problem for Thue
systems \cite[Lemma II]{Post_1947}. Clearly, any original run of $N$ is present
in $M$. Conversely, consider an accepting run in $M$, which may use the new
backward edges. Because there is no outgoing forward edge from the accepting
state, the run cannot end on a reverse edge. Therefore, as long as it contains
reverse edges, there exists a reverse edge $t_1$, which is followed by a
forward edge $t_2$. Observe that $t_1$ can either increment, decrement, or
zero-test the counter $c_i$, for some $i \in \{1,2\}$. Since $N$ is
deterministic, we know that $t_2$ can either decrement, increment, or zero-test
the counter $c_i$, respectively. Hence, the effects of $t_1$ and $t_2$ cancel
out and we can omit both from the run. We iterate this argument and finally
obtain a run without any reverse edges, which is also a run of $N$.

\subparagraph*{The reduction to regular separability}
Given a word $w\in\{0,1\}^*$, we consider $k=2^{n(w)}$. We shall construct two
alphabetically bidirected OCA $V_1$ and $V_2$ such that 
\begin{enumerate}
    \item if $k \in L(M)$ then $L(V_1) \cap L(V_2) \neq \emptyset$ (and thus they are not regularly separable); \label{cond:one}
    \item if $k \notin L(M)$ then $L(V_1)$ and $L(V_2)$ are regularly separable.\label{cond:two}
\end{enumerate}
As the alphabet of $V_1$ and $V_2$, we take $\Sigma = T \cup \overline{T}$, i.e., the transitions of $M$ where each transition $t$ coming from $N$ is represented by the letter $t$ and its added reverse transition by the letter $\overline{t}$. Intuitively, $V_1$ will ignore the second counter of $M$, and $V_2$ will ignore the first counter of $M$. Formally, for $j \in \set{1,2}$, $V_j$ consists of states of $M$ (and one additional state $q_i'$ in the case of $V_1$), and the transition set is constructed as follows. Let $t$ be a transition from $p$ to $q$ in $M$. If $t$ increments, decrements, zero-tests, or performs no operation on the counter $c_j$, then we add the corresponding transition $(p,t,1,q)$, $(p,t,-1,q)$, $(p,t,=0,q)$, or $(p,t,0,q)$, respectively, to $V_j$. Furthermore, to have essentially $q_i(k)$ as the initial configuration in $V_1$, we add transitions $(q_i', \epsilon, k, q_i)$ and $(q_i, \epsilon, -k, q_i')$ to $V_1$.
Observe that $V_j$ is an alphabetically bidirected OCA, because we construct it from $M$, and hence existence of transition $(p,t,\myvec v, q)$ implies existence of the transition $(p,\overline{t}, -\myvec v, q)$. Finally, we set $q_i'$ and $q_i$ to be the initial states of $V_1$ and $V_2$, respectively, and $q_a$ to be the final state in both $V_1$ and $V_2$.

Observe that if we allow binary representations of updates in the OCA (which of
course, does not affect decidability), then this reduction above can indeed be
performed in time exponential in $|w|$, since the $k$-labeled transitions carry
a binary representation with $n(w)\in\Theta(2^{|w|})$ bits.

\subparagraph*{Intersection} Let us show that $V_1$ and $V_2$ satisfy condition~\ref{cond:one}. Let $\rho$ be the shortest run of $M$ which accepts $k$. Hence $\rho$ does not use reverse transitions of $M$. Observe, that after reading word $\rho$ we know that $V_1$ and $V_2$ are both in configuration $q_a(0)$. Hence, $\rho \in L(V_1) \cap L(V_2)$.

\subparagraph*{Regular separability} We now show that $V_1$ and $V_2$ satisfy condition~\ref{cond:two}.
We present a proof sketch here, with full details in \cref{app:separability}.
Suppose $k \notin L(M)$, or equivalently $k \notin L(N)$. This means, the unique run $\rho$ of $N$ (when starting with counter value $k$) ends up in a rejecting state after some number $n$ of steps. Since $N$ is reversible, this means all the paths in $M$ must move along this path of $\rho$: They can walk back and forth, but they must stay on this path. We turn this path (including backward edges) into a DFA $\mathcal{A}$. Observe that for any state $q$ of $\mathcal{A}$ and any label $a$ not leaving $q$, there is at most one of the languages $L(V_1)$ and $L(V_2)$ that can use this label to continue a run: Otherwise we would have a run of $M$ that leaves the path $\rho$. Therefore, for each pair $(q,a)$, we can classify the input into either (potentially) belonging to $L(V_1)$ or (potentially) belonging to $L(V_2)$, but not both. This allows us to construct a regular separator for $L(V_1)$ and $L(V_2)$.

	\section{Conclusion} Our results draw a relatively complete picture of
	decidability and complexity of regular separability in alphabetically
	bidirected systems, even beyond VASS: Undecidability for OCA also implies
	undecidability for alphabetically bidirected pushdown systems. For
	alphabetically bidirected VASS in fixed dimension, \cref{lem:vass_equivalence}
	also implies that the complexity coincides with that of reachability. The
	latter is $\mathsf{L}$-complete under unary
	updates~\cite[Thm.~3.2(1)]{DBLP:conf/lics/GanardiMZ22}.

	Towards resolving the question of decidability of regular separability in the
	``CFL vs.\ VASS'' setting mentioned in the introduction, it would be
	interesting to study regular separability of a pushdown automaton and a VASS,
	where both systems are alphabetically bidirected. Here, the techniques
	for decidability of bidirected pushdown VASS, shown in
	\cite{DBLP:conf/icalp/GanardiMPSZ22}, might be useful.
	
	\label{beforebibliography}
	\newoutputstream{pages}
	\openoutputfile{main.pages.ctr}{pages}
	\addtostream{pages}{\getpagerefnumber{beforebibliography}}
	\closeoutputstream{pages}
	\bibliography{citat}
	\newpage
	\crefalias{section}{appsec}
	\crefalias{subsection}{appsec}
	
	\appendix
	\section{Missing proofs of Section~\ref{sec:attempt}}\label{app:attempt}

\begin{proof}[Proof of Claim~\ref{clm:product}]
    Let $V_1$ and $V_2$ be alphabetically bidirected VASSs over $\Sigma$. We show that $V_1 \times V_2$ is an alphabetically bidirected VASS over $\Sigma$. Let us take any transition $t = ((p_1,p_2), a, (\myvec v_1, \myvec v_2), (q_1,q_2))$ from $V_1 \times V_2$, which goes from state $(p_1,p_2)$ to state $(q_1,q_2)$ over letter $a \in \Sigma$ and has effect $(\myvec v_1, \myvec v_2)$. Because of the definition of a product VASS, we know that for each $i \in [1,2]$ there exists a transition $t_i = (p_i,a,\myvec v_i, q_i)$ in $V_i$. Because both $V_1$ and $V_2$ are alphabetically bidirected VASSs over $\Sigma$ we know that for each $i \in [1,2]$ there exists a transition $t_i'=(q_i, \overline{a},-\myvec v_i, p_i)$ in $V_i$. Hence, there exists a transition $t'= ((q_1,q_2), \overline{a}, (-\myvec v_1, -\myvec v_2), (p_1, p_2)))$ in $V_1 \times V_2$, which is the reverse transition for the transtion $t$. Therefore, $V_1 \times V_2$ is an alphabetically bidirected VASS over $\Sigma$. 
\end{proof}

\section{Missing proofs of Section~\ref{sec:zvass}}
\subsection{Proof of Lemma~\ref{lem:det_zvass}}\label{app:det_zvass}
Here, we show:
\detZVASS*
To prove Lemma~\ref{lem:det_zvass}, we need the following auxiliary lemma.
\begin{lemma}\label{lem:alphabetic morphism}
    Let $K, L \subseteq \Sigma^*$ be two languages and $h: \Gamma^*\rightarrow \Sigma^*$ be an alphabetic morphism\footnote{A morphism $h: \Gamma^* \rightarrow \Sigma^*$ is alphabetic if for each letter $a \in \Gamma$ it holds that $|h(a)| \leq 1$}. If $K' \subseteq h^{-1}(K)$ and $h(K') = K$, then we have:
    \begin{enumerate}
        \item $K \regsepop L \iff K' \regsepop h^{-1}(L)$
        \item $K \cap L = \emptyset \iff K' \cap h^{-1}(L) = \emptyset$
    \end{enumerate}
\end{lemma}
The proof that $K \regsepop L \iff K' \regsepop h^{-1}(L)$ in
\Cref{lem:alphabetic morphism} can be found in the full version of
\cite{new_zvass}, i.e., \cite[Lemma~C.1]{new_zvass_arxiv}. We present a proof
here for completeness, together with the proof of the second equivalence.
\begin{proof}[Proof of Lemma~\ref{lem:alphabetic morphism}]
    Firstly, we prove the following:
    \begin{equation*}
        K \regsepop L \iff K' \regsepop h^{-1}(L)
    \end{equation*}

    We begin with showing, that $K \regsepop L$ implies $K' \regsepop h^{-1}(L)$. Let $R$ be a regular language satisfying $K \subseteq R$ and $L \cap R = \emptyset$. Set $R' = h^{-1}(R)$ and observe that it is a regular language, as regular languages are closed under inverse alphabetic morphisms. Observe that:
    \begin{equation*}
        h^{-1}(L) \cap h^{-1}(R) = h^{-1}(L \cap R) = \emptyset
    \end{equation*}
    And
    \begin{equation*}
        K' \subseteq h^{-1}(K) \subseteq h^{-1}(R) = R' 
    \end{equation*}
    Hence $K' \regsepop h^{-1}(L)$.

    Now, we show that $K' \regsepop h^{-1}(L)$ implies $K \regsepop L$. Let $R$ be a regular language satisfying $K' \subseteq R$ and $h^{-1}(L) \cap R = \emptyset$. Set $R' = h(R)$, which is clearly regular and observe:
    \begin{equation*}
        K = h(K') \subseteq h(R) = R'
    \end{equation*}
    Now observe, that $L \cap R' = \emptyset$. Assume, towards contradiction, that there exists $w \in L \cap R'$. Hence, there exists a word $w' \in R$ such that $h(w') = w$. Moreover, $w' \in h^{-1}(L)$, but this contradicts with $h^{-1}(L) \cap R = \emptyset$. Hence $L \cap R' = \emptyset$ and $K \regsepop L$.

    Now we prove the second part of the lemma:
    \begin{equation*}
        K \cap L = \emptyset \iff K' \cap h^{-1}(L) = \emptyset
    \end{equation*}
    Let $w \in K \cap L$ then, as $K = h(K')$ there exists $w' \in K'$ such that $h(w') = w$. Therefore as $w \in L$ we have also $w' \in h^{-1}(L)$.

    On the other hand let $w \in K' \cap h^{-1}(L)$ then $h(w) \in K \cap L$ (as $h(K') = K$ and $w \in h^{-1}(L)$). Therefore 
    \begin{equation*}
        K \cap L = \emptyset \iff K' \cap h^{-1}(L) = \emptyset
    \end{equation*}
    This concludes the proof of Lemma~\ref{lem:alphabetic morphism}.
\end{proof}

 Now we are ready to prove \Cref{lem:det_zvass}.
\begin{proof}[Proof of Lemma~\ref{lem:det_zvass}]
    The proof of this lemma is similar to \cite[Lemma~7]{z_vass_separability_classical}, but we work in the bidirected setting. For $i \in \set{1,2}$ let $V_i = (Q_i, \Sigma, T_i, q_i, p_i)$ be an alphabetically bidirected \zvass over $\Sigma = \Sigma' \cup \overline{\Sigma'}$. From $V_1$ and $V_2$ we construct two alphabetically bidirected \zvass $V_1'$  and $V_2'$ over  $\Gamma$ such that:
    \begin{equation*}
        L(V_1) \regsepop L(V_2) \iff L(V_1') \regsepop L(V_2')
    \end{equation*}
    And:
    \begin{equation*}
    L(V_1) \cap L(V_2) = \emptyset \iff L(V_1') \cap L(V_2') = \emptyset
    \end{equation*}
    Moreover, we want $V_1'$ to be deterministic. We obtain $V_1'$ by making each label of a transition in $V_1$ unique. Concretely, let $T$ be the set of transitions over letters from $\Sigma'$ in $V_1$. Then for each $t \in T$ we replace its label by $t$. For all other transitions, namely for each transition $t' \in T_1 \setminus T$, there exists a transition $t \in T$, which is the reverse transition of $t'$, as $V_1$ is an alphabetically bidirected VASS. We replace the label of $t'$ by $\overline{t}$. Hence, clearly $V_1'$ is an alphabetically bidirected deterministic \zvass over $\Gamma = T \cup \overline{T}$. From the above construction we also get a morphism $h: \Gamma^* \rightarrow \Sigma^*$, which is induced by replacing of labels of transitions. Concretely, if a transition with label $a \in \Sigma_ \epsilon$  was replaced by a transition with label $b \in \Gamma$, then $h(b) = a$. The morphism $h$ is naturally extended to words. We observe that the following claim holds, which will allow us to apply Lemma~\ref{lem:alphabetic morphism} later. The proof of \Cref{clm:mor_V1} is provided after this proof.
    \begin{claim}\label{clm:mor_V1}
        Languages $L(V_1)$ and $L(V_1')$ and morphism $h$ satisfy the following properties:
        \begin{enumerate}
            \item $L(V_1') \subseteq h^{-1}(L(V_1))$
            \item $h(L(V_1')) = L(V_1)$
        \end{enumerate}
    \end{claim}

    Now we obtain $V_2'$ from $V_2$ using the morphism $h$ as follows. We replace each transition $t \in V_2$ with many transitions, all with the same effect and between the same states as $t$,  one for each label $a \in \Gamma$ such that $h(a) = b$, where $b$ is the label of $t$. Additionally, we add loops labelled with $a \in \Gamma$ such that $h(a) = \epsilon$ to any state of $V_2'$. 
    
    It is easy to observe, that $L(V_2') = h^{-1}(L(V_2))$ and if $V_2$ was deterministic $\epsilon$-free than $V_2'$ is also deterministic. Additionally, $V_1'$ and $V_2'$ can be constructed in polynomial time, and they are alphabetically bidirected \zvass over $\Gamma$.

    Now, because of Claim~\ref{clm:mor_V1}, equality $L(V_2') = h^{-1}(L(V_2))$ and the fact that $h$ is an alphabetic morphism, using Lemma~\ref{lem:alphabetic morphism} we obtain:
    \begin{equation*}
        L(V_1) \regsepop L(V_2) \iff L(V_1') \regsepop L(V_2')
    \end{equation*}
    And:
    \begin{equation*}
    L(V_1) \cap L(V_2) = \emptyset \iff L(V_1') \cap L(V_2') = \emptyset
    \end{equation*}

    Finally, we can apply the same polynomial-time procedure to $V_2'$ and $V_1'$ to determinize $V_2'$. The result is two alphabetically bidirected \zvass $W_2$ (obtained from $V_2'$) and $W_1$ (obtained from $V_1'$). It is clear by construction that $W_2$ is deterministic. Moreover, also $W_1$ is deterministic as $V_1'$ is deterministic $\epsilon$-free. Hence, we get:
    \begin{equation*}
        L(V_1) \regsepop L(V_2) \iff L(W_1) \regsepop L(W_2)
    \end{equation*}
    And:
    \begin{equation*}
    L(V_1) \cap L(V_2) = \emptyset \iff L(W_1) \cap L(W_2) = \emptyset
    \end{equation*}
    This concludes the reduction and proof of Lemma~\ref{lem:det_zvass}.
\end{proof}

\begin{proof}[Proof of Claim~\ref{clm:mor_V1}]
        Firstly, we show that $L(V_1') \subseteq h^{-1}(L(V_1))$.  Let $w \in L(V_1')$. We aim at showing $w \in h^{-1}(L(V_1))$. Observe that there is an accepting run $t_1't_2'\ldots t_n'$ in $V_1'$ with $t_i' = (q_i, t_i, \myvec x_i, q_{i+1})$. Notice, that $t_i \in \Gamma = T \cup \overline{T}$. Hence, they are transitions of $V_1$. Observe that for each $i \in [1,n]$ it holds that $t_i = (q_i, a_i, \myvec x_i, q_{i+1})$ for some $a_i \in \Sigma_\epsilon$. Therefore $a_1a_2\ldots a_n \in L(V_1)$. Moreover, $h(w) = h(t_1t_2\ldots t_n) = a_1a_2\ldots a_n$, which implies $w \in h^{-1}(L(V_1))$.

        Now we prove that $h(L(V_1')) = L(V_1)$. A word $w \in \Sigma^*$ is in $h(L(V_1'))$ if, and only if, there is a word $w' \in L(V_1')$ such that $w = h(w')$. Let $t_1't_2'\ldots t_n'$ be an accepting run in $V_1'$, that is labelled with $w'$. Hence, for each $i \in [1,n]$, we have $t_i' = (q_i, t_i, \myvec x_i, q_{i+1})$ and $w' = t_1t_2\ldots t_n$. By construction this is equivalent to an accepting run $t_1t_2\ldots t_n$ in $V_1$ over $h(t_1)h(t_2)\ldots h(t_n) = h(t_1t_2 \ldots t_n) = h(w') = w$. Hence, equivalent to $w \in L(V_1)$.
\end{proof}

\subsection{Proof of Lemma~\ref{lem:one_zvass}}\label{app:one_zvass}
\begin{proof}[Proof of \Cref{lem:one_zvass}]
    Let $V_1$ and $V_2$ be alphabetically deterministic bidirected \zvass with $d_1$ and $d_2$ counters, respectively. We use product construction and obtain the deterministic $(d_1+d_2)$-\zvass $V = V_1 \times V_2$, which is an alphabetically bidirected \zvass because of Claim~\ref{clm:product}. We set $I_1 = [1, d_1]$ and $I_2 = [d_1 + 1 , d_1, + d_2]$ and show, that $L(V_1) \regsepop L(V_2)$ holds if, and only if, $L(V, I_1) \regsepop L(V, I_2)$ holds. Let $A_i$ be the underlying DFA of $V_i$, namely $V_i$ with removed counter updates. Observe, that $L(V, I_1) = L(V_1) \cap L(A_2)$ and $L(V, I_2) = L(V_2) \cap L(A_1)$.

    Firstly, we show that $L(V_1) \regsepop L(V_2)$ implies $L(V, I_1) \regsepop L(V, I_2)$. Observe, that then there exists a regular language $R \subseteq \Sigma^*$ such that $L(V_1) \subseteq R$ and $L(V_2) \cap R = \emptyset$. Therefore we have 
    $$L(V, I_1) = L(V_1) \cap L(A_2) \subseteq L(V_1) \subseteq R $$
    and 
    $$L(V, I_2) \cap R = (L(V_2) \cap R) \cap L(A_1)   = \emptyset$$
    Hence $L(V, I_1) \regsepop L(V, I_2)$ holds.

    Now, we show the opposite direction. Because $L(V, I_1) \regsepop L(V, I_2)$ there exists a regular language $R \subseteq \Sigma^*$ such that $L(V, I_1) \subseteq R$ and $L(V, I_2) \cap R = \emptyset$. Let us set $R' = (R \cap L(A_1)) \cup (\Sigma^* \setminus L(A_2))$, which is clearly a regular language. Observe that:
    \begin{align*}
    L(V_1) &\subseteq (L(V_1) \cap L(A_2)) \cup (\Sigma^* \setminus L(A_2)) \\
    &= (L(V, I_1) \cap L(A_1)) \cup (\Sigma^* \setminus L(A_2)) \\
    &\subseteq (R \cap L(A_1)) \cup (\Sigma^* \setminus L(A_2)) = R'
    \end{align*}
    On the other hand:
    \begin{align*}
    L(V_2) \cap R' &= (L(V_2) \cap L(A_1) \cap R) \cup [L(V_2) \cap (\Sigma^* \setminus L(A_2))]\\ &\subseteq (L(V, I_1) \cap R) \cup [L(A_2) \cap (\Sigma^* \setminus L(A_2))] = \emptyset
    \end{align*}
    Hence $L(V_2) \cap R' = \emptyset$ and $L(V_1) \regsepop L(V_2)$ holds.

    Finally, we need to show, that $L(V_1) \cap L(V_2) = L(V, I_1) \cap L(V, I_2)$ Observe, that:
    $$L(V_1) \cap L(V_2) = L(V) \subseteq L(V, I_1) \cap L(V, I_2)$$
    On the other hand:
    $$L(V, I_1) \cap L(V, I_2) = L(V_1) \cap L(A_2) \cap L(V_2) \cap L(A_1) \subseteq L(V_1) \cap L(V_2)$$
    Hence $L(V_1) \cap L(V_2) = L(V, I_1) \cap L(V, I_2)$. This concludes the proof of Lemma~\ref{lem:one_zvass}.
\end{proof}

\subsection{Definition of skeletons}\label{app:skeletons}
We define $\skel$ inductively on the length of the path $\pi$. Namely, we set $\skel(\epsilon) = \epsilon$ and for a path  $\pi=t_1t_2\ldots t_n$ we consider two cases. Assume, that $\skel(t_1t_2\ldots t_{n-1}) = s_1s_2\ldots s_j$ and let $q$ be the state in which transition $t_n$ ends. 

The first case is that there is no $i \in [1,j]$ such that the transition $s_i$ ends in the state $q$. Then we set $\skel(\pi) = s_1s_2\ldots s_jt_n$.
Observe that because by inductive definition, $s_1s_2\ldots s_j$ ends in the same state as $t_{n-1}$ and is a path, we have that $\skel(\pi)$ is a path ending in the same state as $\pi$.

The second case is that there exists $i \in [1,j]$ such that $s_i$ ends in $q$. Let us choose a maximal such $i$. Now observe, that as $s_1s_2\ldots s_j$ is a path we have that $s_{i+1}s_{i+2}\ldots s_j t_n$ is a cycle in $V$.

If all states occurring in the cycle also occur in the path $s_1s_2\ldots s_i$, then we set $\skel(\pi) = s_1s_2\ldots s_i$, i.e., we omit the cycle in the skeleton as it does not increase the set of visited states. Note that $s_1s_2\ldots s_i$ is a
path ending in $q$, that is the same state as path $\pi$.

Otherwise, at least one state in the cycle $s_{i+1}s_{i+2}\ldots s_j t_n$ does not occur in the path $s_1s_2\ldots s_i$. In this case, we set $\skel(\pi) =  s_1s_2\ldots s_jt_n$ and therefore $\skel(\pi)$ is also a path ending in the state $q$, namely the same state as the end state of $\pi$. 

\subsection{Proof of Lemma~\ref{lem:zvass_equivalence}}\label{app:zvass_equivalence}
Here, we prove the following:
\zvassEquivalence*
\begin{proof}
We know, that $L(V_1) \cap L(V_2) = \emptyset$ and want to prove, that $L(V_1) \regsepop L(V_2)$. Firstly, we use Lemma~\ref{lem:det_zvass} and obtain deterministic alphabetically bidirected \zvass $W_1$ and $W_2$ such that $L(W_1) \cap L(W_2) = \emptyset$. Moreover, because of Lemma~\ref{lem:det_zvass} it is enough to prove $L(W_1) \regsepop L(W_2)$. Now, we use Lemma~\ref{lem:one_zvass} and obtain deterministic alphabetically bidirected VASS $V$ and two sets $I_1, I_2$ such that it is enough to prove $L(V, I_1) \regsepop L(V, I_2)$ and moreover $L(V, I_1) \cap L(V, I_2) = L(W_1) \cap L(W_2) = \emptyset$. Let $\rho$ be a skeleton. Because of Lemma~\ref{lem:zvass_skeleton} it is enough to prove $L(V, I_1, \rho) \regsepop L(V, I_2, \rho)$. Moreover, by definition $L(V, I_1, \rho) \subseteq L(V, I_1)$ and hence we have $L(V, I_1, \rho) \cap L(V, I_2, \rho) = \emptyset$. Finally, by Lemma~\ref{lem:zvass_sets}, it is enough to prove $M(V, I_1, \rho) \regsepop M(V, I_2, \rho)$. Observe, that because $L(V, I_1, \rho) \cap L(V, I_2, \rho) = \emptyset$ we have $M(V, I_1, \rho) \cap M(V, I_2, \rho) = \emptyset$. By Claim~\ref{clm:hyperlinear} we know, that $M(V, I_1, \rho) = A + U^*$ and $M(V, I_2, \rho) = B + V^*$ for some finite sets $A,B, U, V$. By Lemma~\ref{lem:sep_characterization} it is enough to show, that $(A + U^* + V_J^*) \cap (B + V^* + U_J^*) = \emptyset$ where $J$ is the set of coordinates that are twin-unbounded for $M(V, I_1, \rho), M(V, I_2, \rho)$. Assume, towards contradiction, that $(A + U^* + V_J^*) \cap (B + V^* + U_J^*) \neq \emptyset$. That means there exists $\myvec a \in A$, $\myvec u \in U^*$, $\myvec v_J \in V_J^*$ and $\myvec b \in B$, $\myvec v \in V^*$, $\myvec u_J \in U_J^*$ such that $\myvec a + \myvec u + \myvec v_J = \myvec b + \myvec v + \myvec u_J$. Hence $\myvec a + \myvec u - \myvec u_J = \myvec b + \myvec v - \myvec v_J$. Observe that $\myvec a + \myvec u-\myvec u_J \in \Z^{|S|}$ (i.e., not every component needs to be nonnegative). We want to find a vector $\myvec w$ such that $\myvec a + \myvec u - \myvec u_J + \myvec w \in M(V, I_1, \rho) \cap M(V, I_2, \rho)$, which will give us contradiction with $M(V, I_1, \rho) \cap M(V, I_2, \rho) = \emptyset$. For that, we need the following claim where $\1_{\pi}$ means a vector in $\N^{|S|}$ where all entries are equal to zero except the one corresponding to cycle $\pi$, which is equal to $1$; the claim is proved in the appendix.
\begin{claim}\label{clm:every_cycle}
    Let $\pi$ be $\rho$-cycle. Then $\1_{\pi} + \1_{\overline{\pi}} \in U^* \cap V^*$. 
\end{claim}

Let $n \in \N$ be such that for each $i \in [1, |S|]$ we have $|(\myvec a + \myvec u - \myvec u_J)_i| \leq n$. We define vector $\myvec w = \sum_{\pi \in S} n(\1_{\pi} + \1_{\overline{\pi}}) \in U^* \cap V^*$ (by Claim~\ref{clm:every_cycle}). Now observe, that $\myvec a + \myvec u - \myvec u_J + \myvec w \in \N^{|S|}$. Now observe, that 
$$\eff_{I_1}(\rho) + \eff_{I_1}(\myvec a+\myvec u-\myvec u_J+\myvec w) = \eff_{I_1}(\rho) + \eff_{I_1}(\myvec a)$$ and hence $\myvec a + \myvec u - \myvec u_J+\myvec w \in M(V, I_1, \rho)$. On the other hand, 
$$\eff_{I_2}(\rho) + \eff_{I_2}(\myvec a+\myvec u-\myvec u_J+\myvec w) = \eff_{I_2}(\rho) + \eff_{I_2}(\myvec b+\myvec v-\myvec v_J+\myvec w) = \eff_{I_2}(\rho) + \eff_{I_2}(\myvec b)$$ and hence $M(V, I_2, \rho)$.  Therefore, $\myvec a + \myvec u - \myvec u_J + \myvec w \in M(V, I_1, \rho) \cap M(V, I_2, \rho)$, which gives us a contradiction with $M(V, I_1, \rho) \cap M(V, I_2, \rho) = \emptyset$. Hence $(A + U^* + V_J^*) \cap (B + V^* + U_J^*) = \emptyset$, which finishes the proof, that $L(V_1) \regsepop L(V_2)$ and concludes the proof of Lemma~\ref{lem:zvass_equivalence}.
\end{proof}

\begin{proof}[Proof of \Cref{clm:every_cycle}]
    Let $\myvec v = \1_{\pi} + \1_{\overline{\pi}}$. Observe, that $\overline{\pi}$ is also a $\rho$-cycle. Moreover, we have $\eff(\pi) + \eff(\overline{\pi}) = \vec 0$. Hence, if $\myvec u \in M(V, I_i, \rho)$ then also $\myvec u + \myvec v \in M(V, I_i, \rho)$. Therefore $\myvec v \in U^* \cap V^*$. Note, that it is not always true, that $\myvec v \in U \cap V$ as it can for instance happen, that $\set{\1_{\pi}, \1_{\overline{\pi}}} \subseteq U$, but $\myvec v \notin U$.  
\end{proof}

\section{Missing proofs of Section~\ref{sec:vass}}\label{app:vass}
\subsection{Hardness proof}\label{app:hardness}

	We reduce from reachability in bidirected VASS, which is
	$\expspace$-complete~\cite[Main Theorem]{mayr1982complexity}.

	Given bidirected VASS $V$, source configuration $q_i(\vec{0})$ and target configuration $q_f(\vec{0})$. We construct an alphabetically bidirected VASS $V_1$. $V_1$ has the same structure of underlying graph as $V$, initial state $q_i$ and accepting state $q_f$. We split all transitions of $V_1$ into $n$ pairs $(t_{i_1}, t_{i_2})$ in such a way that each transition is paired only with its reverse transition. For each pair $(t_{i_1}, t_{i_2})$ we label the transition $t_{i_1}$ with letter $t_{i_1}$ and the transition $t_{i_1}$ with letter $\overline{t_{i_1}}$. Observe that the constructed VASS $V_1$ is an alphabetically bidirected VASS over $\Sigma = \Sigma_1 \cup \overline{\Sigma_1}$, where $\Sigma_1 = \bigcup_{i \in [1,n]}\set{t_{i_1}}$. This is because, for each $i \in [1,n]$, if a transition is over $t_{i_1}$ then there exists a reverse transition over $\overline{t_{i_1}}$ and vice versa. Moreover, we set $V_2$ to be a single state with transitions over each letter $a \in \Sigma$, which is clearly an alphabetically bidirected VASS over $\Sigma$. Notice that non-reachability of $q_f(\vec{0})$ from $q_i(\vec{0})$ in $V$ is equivalent to $L(V_1) \cap L(V_2) = \emptyset$, which, because of Lemma \ref{lem:vass_equivalence}, is equivalent to regular separability of $L(V_1)$ and $L(V_2)$. Hence, regular separability of languages of bidirected VASSs over bidirected alphabet is \expspacehard.

\subsection{Proof of Claim~\ref{clm:reverse}}\label{app:vass_1}
\begin{proof}[Proof of \Cref{clm:reverse}]
    As $\rev(\rev(L_i)) = L_i$ it is enough to prove implication that if $L_1 \regsepop L_2$ then $\rev(L_1) \regsepop \rev(L_2)$.

    Let $R$ be a regular separator of $L_1$ and $L_2$ over alphabet $\Sigma$. We claim that the following language $R' = \setof{\overline{w}}{w \in R}$ is a regular separator of $\rev(L_1)$ and $\rev(L_2)$. Observe, that $R'$ is a regular language as regular language are closed under reversing words and if $L$ is a regular language than $\setof{\overline{w_1}\overline{w_{2}}\ldots \overline{w_n}}{w_1w_2\ldots w_n \in L}$ is also a regular language as it is enough to change label of each transition in automata recognising $L$ from $a$ to $\overline{a}$. Hence, we only need to prove, that $R'$ separates $\rev(L_1)$ and $\rev(L_2)$.

    Firstly, observe that $\rev(L_1) \subseteq R'$. Indeed, for each word $w \in \rev(L_1)$ we have that $\overline{w} \in L(V_1) \subseteq R$. Hence $w = \overline{u}$, for $u \in R$, and therefore $w \in R'$.

    Moreover, observe, that $\rev(L_2) \cap R' = \emptyset$. Assume, towards contradiction, that there exists $w \in \rev(L_2) \cap R'$. Then there exists $u \in R$ such that $w = \overline{u}$. Moreover, $u = \overline{w} \in L_2$, but this contradicts with the fact that $L_2 \cap R = \emptyset$.
\end{proof}

\subsection{Proof of Claim~\ref{clm:covering_word}}
\begin{proof}[Proof of \Cref{clm:covering_word}]\label{app:vass_2}
    We only show the existence of $u_1$, relying on the fact that for $u_2$ the proof is symmetric. Recall that $(V_1, V_2)$ was simultaneously unbounded and since we get a run $\rho$ in $V_1 \times V_2$ from $q(\vec{0})$ to $p(\myvec v)$ such that $\myvec v > \vec{1}$, where $q$ and $p$ are, respectively, the initial and the final state of $V_1 \times V_2$. For $i \in [1,2]$, let $\myvec v_i$ be $\myvec v$ projected to the counters from the VASS $V_i$. Let us set the word read by $\rho$ as $u_1$. For $j \in [1,2]$, the run $\rho$ yields a run $\rho_{u_1}^j$ from $q_j(\vec{0})$ to $p_j(\myvec v_i)$ in $V_j$ where $q_j$ and $p_j$ are, respectively, the initial and the final state of $V_j$. Since $\myvec v_i > \vec{1}$ we get that it covers $p_i(\vec{1})$. Moreover, $\eff(\rho_{u_1}^i) = \myvec v_i  \geq \vec{1}$. Hence $u_1$ satisfies conditions of Claim~\ref{clm:covering_word}.
\end{proof}

\subsection{Proof of Lemma~\ref{lem:vass_equivalence}}\label{app:vass_3}
\begin{proof}[Proof of \Cref{lem:vass_equivalence}]
The only non-obvious direction is the implication that intersection non-emptiness implies regular separability. We proceed by induction on $d_1+d_2$. If $d_1 + d_2 = 0$ then the property of Lemma~\ref{lem:vass_equivalence} is clearly true because $L(V_1)$ and $L(V_2)$ are regular languages. Now assume that $d_1 + d_2 > 0$. If $(V_1, V_2)$ and $(\rev(V_1), \rev(V_2))$ are simultaneously unbounded then, by \Cref{clm:vass_regsep_easy}, we get $L(V_1) \regsepop L(V_2)$, as required. Otherwise, because of \Cref{clm:reverse}, we can w.l.o.g. assume that $(V_1, V_2)$ is not simultaneously unbounded. Let us consider VASS $V_1 \times V_2$, which is an alphabetically bidirected VASS due to Claim~\ref{clm:product} and partition its counters into two sets $P_1$ and $P_2$ as in Lemma~\ref{lem:partition}. Since $(V_1, V_2)$ is not simultaneously unbounded we get that $P_1 \neq \emptyset$. \Wlog assume that there exists counter $i$ coming from $V_1$ such that $i \in P_1$. Let $V_{1,\leq B}$ be the VASS constructed from $V_1$ by tracking counter $i$ up to $B$ (not allowing it to exceed $B$). Concretely, let $V_1 = (\Sigma, Q_1, T_1,q_i, q_f)$ then $V_{1,\leq B} = (\Sigma, Q_1', T_1', q_i^0, q_f^0)$ where $$Q_1' = \setof{q^b}{q \in Q_1, b \in [0,B]}$$
and 
$$T_1' = \setof{(q^b,a, \pi_{-i}(\myvec v), p^{b'})}{(q,a,\myvec v, p) \in T_1, b + \myvec v_i = b', b, b' \in [0,B]}$$
Observe that $V_{1,\leq B}$ is an alphabetically bidirected VASS. Indeed, the existence of a transition $(q^b,a,\myvec u,p^{b'})$ in $V_{1,\leq B}$ implies the existence of the transition $(q,a,\myvec v,p)$, for some $\myvec v \in \Z^{d_1}$ such that $\pi_{-i}(\myvec v) = \myvec u$, in $V_1$. Since $V_1$ is alphabetically bidirected, the transition $(p,\overline{a},-\myvec v,q)$ also exists in $V_1$, which, in turn, implies the existence of the transition $(p^{b'},\overline{a},-\myvec u,q^b)$ in $V_{1,\leq B}$.

Let also $V_{1,>B}$ be a VASS tracking counter $i$ in the states up to the point when it exceeds $B$, then it goes to a special state $q$. To make VASS $V_{1,>B}$ bidirected for each transition incoming to $q$, we add a reverse transition. Moreover, for technical reasons, for each counter $i$ and each letter $a \in \Sigma_\epsilon$, we add a transition $(q, a, -\myvec e_i, q)$ and reverse transitions, i.e.,  $(q, \overline{a}, \myvec e_i, q)$, incrementing the counter (to preserve bidirectedness). We also set the initial state of $V_{1,>B}$ to the initial state of $V_1$ (with counter $i$ put into states) and final state to $q$. Concretely, $V_{1,>B} = (\Sigma, Q_1' \cup \set{q}, T_1' \cup T_1'', q_i^0, q)$ where
\begin{align*}
T_1'' &= \setof{(q^b, a, \pi_{-i}(\myvec v), q)}{(q, a, \myvec v, p) \in T_1, b + \myvec v_i > B} \\ &\cup \setof{(q, \overline{a}, -\pi_{-i}(\myvec v), q^b)}{(q, a, \myvec v, p) \in T_1, b + \myvec v_i > B} \\ &\cup \setof{(q, a, -\myvec e_i, q)}{i \in [d_1], a \in \Sigma_\epsilon} \cup \setof{(q, \overline{a}, \myvec e_i, q)}{i \in [d_1], a \in \Sigma_\epsilon}
\end{align*}

 We observe that the following claim holds, which  allows us to apply the induction assumption. The proof of \Cref{clm:good_for_induction} is provided after this proof.
\begin{claim}\label{clm:good_for_induction}
    The following properties hold:
    \begin{itemize}
        \item $L(V_1) \subseteq L(V_{1,\leq B}) \cup L(V_{1,>B})$
        \item $(L(V_{1,\leq B}) \cup L(V_{1,>B})) \cap L(V_2) = \emptyset$
    \end{itemize}
\end{claim}
Because of Claim~\ref{clm:good_for_induction}, we can apply the induction assumption for $V_{1,\leq B}$ and $V_2$ and get a regular separator $R_1$. Similarly, we can apply the induction assumption for $V_{1,>B}$ and $V_2$ and get a regular separator $R_2$. Hence $R_1 \cup R_2$ is a regular separator of $L(V_1)$ and $L(V_2)$. This finishes the proof of \Cref{lem:vass_equivalence}.
\end{proof}

\begin{proof}[Proof of \Cref{clm:good_for_induction}]
    Let $w \in L(V_1)$ and let $\rho$ be a run accepting $w$. One case is that in every configuration $p(\myvec u)$ in $\rho$, we have that $\myvec u_i \leq B$. Observe that since $V_{1,\leq B}$ is a VASS constructed from $V_1$ by tracking counter $i$ in the states up to value $B$, we can conclude that $w \in L(V_{1,\leq B}) \subseteq L(V_{1,\leq B}) \cup L(V_{1,>B})$. The second case is that there exists a configuration $p(\myvec u)$ of $\rho$ such that $\myvec u_i > B$. Observe that $V_{1,>B}$ is a VASS constructed from $V_1$ by tracking counter $i$ in a state up to the point when it exceeds $B$ and then going to special state $q$ where for each counter and each letter $a \in \Sigma_\epsilon$ we have two self-loops from $q$ to $q$ over $a$, one decrementing the counter and the second incrementing the counter. Observe that after reading the word read by a prefix of $\rho$ ending in configuration $p(\myvec u)$, we can reach state $q$ in $V_{1,>B}$. Then we can read the remaining suffix by transitions, incrementing counters, and reach some configuration $q(\myvec u')$. Finally, we can reach $q(\vec 0)$, which is an accepting configuration in $V_{1,>B}$, using $\epsilon$-transitions and accept $w$. Hence $w \in L(V_{1,>B})$.

    Assume, towards contradiction, that there exists $w \in (L(V_{1,\leq B}) \cup L(V_{1,>B})) \cap L(V_2)$. Observe, that if $w \in L(V_{1,\leq B})$, since $V_{1,\leq B}$ is a VASS constructed from $V_1$ by tracking counter $i$ in the states up to value $B$, we get $w \in L(V_1)$, which contradicts $L(V_1) \cap L(V_2) = \emptyset$. Hence $w \in L(V_{1,>B}) \cap L(V_2)$. Since $V_{1,>B}$ is a VASS constructed from $V_1$ by tracking counter $i$ in a state up to the point when it exceeds $B$ and then going to special state $q$, and the accepting configuration of $V_{1,>B}$ is $q(\vec{0})$, we get that there is a prefix $v$ of $w$ such that it is read by a run $\rho$ in $V_1$ which ends in a configuration where the counter $i$ is bigger than $B$. Since $w \in L(V_2)$, its prefix $v$ is also read by some run in $V_2$ and therefore there exists a run in $V_1 \times V_2$ reading $v$ from the initial configuration and reaching configuration $p(\myvec u)$ with $\myvec u_i > B$, but this contradicts $i \in P_1$. Therefore  $(L(V_{1,\leq B}) \cup L(V_{1,>B})) \cap L(V_2) = \emptyset$ and this concludes the proof of the Claim~\ref{clm:good_for_induction}.
\end{proof}

\section{Missing proofs of Section~\ref{sec:undecidability}}\label{app:undecidability}
\begin{proof}[Proof of \Cref{lem:conterexample}]
    Let us take $V_1$ and $V_2$ presented in the Figure~\ref{fig:conterexample}. Observe that $V_1$ and $V_2$ are alphabetically bidirected OCA. Firstly, we show that $L(V_1)$ and $L(V_2)$ are not regularly separable. Assume, towards contradiction, that they are regularly separable by a language $L$ such that $L(V_1) \subseteq L$ and $L(V_2) \cap L = \emptyset$. Let $N$ be a number bigger than the number of states of the minimal DFA recognising $L$. Observe, that $$t_1^Nt_2t_3^Nt_4t_5^{N!-1} \in L(V_1) \subseteq L$$
    Because of the definition of $N$ there exists $K \leq N$ such that for every $k \in \N$ we have:
    $$t_1^{N+kK}t_2t_3^Nt_4t_5^{N!-1} \in L$$
    Hence $w = t_1^{N+N!}t_2t_3^Nt_4t_5^{N!-1} \in L$, but observe that also $w \in L(V_2)$, which contradicts $L \cap L(V_2) = \emptyset$. Therefore $L(V_1)$ and $L(V_2)$ are not regularly separable.

    Now we show, that $L(V_1) \cap L(V_2) = \emptyset$. Let us take any $w \in L(V_1)$. Observe that $w$ contains the letter $t_4$; let $u$ be the longest prefix of $w$ such that it does not contain $t_4$. Observe that because of the zero-test done by transition over $t_4$, it holds:
    \begin{align}\label{eq:count_letters}
    \countletters{t_1}{u} + \countletters{\overline{t_3}}{u}
    = \countletters{\overline{t_1}}{u} + \countletters{t_3}{u}
    \end{align}

    Assume, towards contradiction, that $w \in L(V_2)$. Now observe that because of the decrement done by the transition over $t_4$, it holds:
    \begin{equation*}\countletters{t_1}{u} + \countletters{\overline{t_3}}{u} > \countletters{\overline{t_1}}{u} + \countletters{t_3}{u}
    \end{equation*}
    But this contradicts Equation~\ref{eq:count_letters}. Hence $w \notin L(V_2)$ and $L(V_1) \cap L(V_2) = \emptyset$.
\end{proof}

\subsection{Proof of regular separability}\label{app:separability}
Here, we prove that for the OCA $V_1$ and $V_2$ in \cref{sec:undecidability}, we have indeed that if $k\notin L(M)$, then $L(V_1)$ and $L(V_2)$ are regularly separable.

Observe, that $L(V_1)$ and $L(V_2)$ are disjoint languages as any word $w \in L(V_1) \cap L(V_2)$ is an accepting run of $M$. Let $n$ be the length of the rejecting run of $N$. We claim that in order to classify a word between $L(V_1)$ and $L(V_2)$, it suffices to inspect prefixes of the word up to the point at which either the number of reverse letters exceeds the number of forward letters or the number of forward letters exceeds the number of the reverse letters by at least $n+1$. For a word $w \in \Sigma^*$ we set
$\phi(w) := \countletters{T}{w} - \countletters{\overline{T}}{w}$.
Now, let us define the language $L$, which consists of words such that for every prefix $v$ we have $\phi(v) \in [0,n]$.
$$L := \setof{w}{w \in \Sigma^*, \minprefix{\phi}{w} \geq 0, \maxprefix{\phi}{w} \leq n}$$

Observe that $L$ is a regular language, because an automaton recognising $L$ needs only to store in its states, after reading prefix $v$, the current value of $\phi(v)$ between $0$ and $n$ and reject as soon as $\phi(v) \notin [0,n]$. For $i \in \set{1,2}$ let $L_i$ contain all words which can be read by $V_i$ from the initial configuration (not necessarily ending in the accepting configuration). Let also
$$K := \bigcup_{a \in \Sigma} \setof{wa}{w \in L_1 \cap L_2 \cap L, [wa \notin L_2 \lor (\phi(wa) \in \set{-1,n+1} \land wa \in L_1)]}$$
We claim that the following language is a regular separator for $L(V_1)$ and $L(V_2)$:
$$S = (L(V_1) \cap L_2 \cap L) \cup K\Sigma^*$$

Firstly, we show that $L(V_1) \subseteq S$. Let us take $w \in L(V_1)$. Let $v$ be the longest prefix of $w$ such that $v \in L_1 \cap L_2 \cap L$ and let $a \in \Sigma_\epsilon$ be the next letter of $w$ after the prefix $v$ or $\epsilon$ if $w = v$. If $a = \epsilon$ we get that $v = w$ and hence $w \in L(V_1) \cap L_2 \cap L \subseteq S$. Otheriwse, as $v$ was the longest prefix of $w$ belonging to $L_1 \cap L_2 \cap L$, we get that $va \notin L_2$ or $va \notin L$ (i.e, $\phi(va) \in \set{-1, n+1}$). Thus, $va \in K$, and hence $w \in S$.

Secondly, we show that $S \cap L(V_2) = \emptyset$. Assume, towards contradiction, that there exists $w \in S \cap L(V_2)$. Since $L(V_1)$ and $L(V_2)$ are disjoint $w \notin L(V_1) \cap L$. Therefore, there exists a prefix $v \in K$ of $w$ such that $\phi(v) \in \set{-1,n+1}$ and $v \in L_1$. Hence $v \in L_1 \cap L_2$. Let $\rho_i$ be a run in $V_i$ that reads $v$. If $v \notin T^* \cup \overline{T}^*$ we can find an infix $t_1t_2$ of $v$ such that $t_1t_2 \in T\overline{T} \cup \overline{T}T$. Since $v$ is read by both $V_1$ and $V_2$, it is also read by $M$. Therefore, similarly to the proof that $L(N) = L(M)$, as $N$ is deterministic and reversible, we can argue that $t_1 = \overline{t_2}$. Hence, the effects of transitions reading $t_1$ and $t_2$ in $\rho_1$ and $\rho_2$ cancel out, and therefore we can remove them from both of the runs and obtain two runs reading the same word $v'$ such that $\phi(v') = \phi(v)$ and $v' \in L_1 \cap L_2$. By repeating this procedure we obtain a word $u \in (T^* \cup \overline{T}^*) \cap L_1 \cap L_2$ and $\phi(u) = \phi(v) \in \set{-1,n+1}$. As the only transition leaving $q_i$ in $V_2$ is over a letter from $T$ we get $\phi(u) = n+1$. However, we cannot have $\phi(u) = n+1$, as then after reading $\prefix{n}{u}$ OCA $V_1$ and $V_2$ will be in configuration $q_r(0)$, and there is no transition over a letter from $T$ leaving the state $q_r$. Hence, $S \cap L(V_2) = \emptyset$.

Hence, what is left is the proof that $S$ is a regular language.
For a word $w \in \Sigma^*$, let us define $\psi(w)$ as the word obtained after repetitive deletion (as long as it is possible) of infixes $\overline{t}t$ or $t\overline{t}$ where $t \in T$. Observe that any word  $w \in L_1 \cap L_2 \cap L$ is a run of $M$ (necesarily starting with transition from $T$). Since $N$ is deterministic and reversible, similarly as in the proof that $L(N) = L(M)$, we can argue that $\psi(w) \in T^*$, is still a run of $M$ (hence $\psi(w) \in L_1 \cap L_2$). Moreover, removing the infix $\overline{t}t$ does not cause a word to not belong to $L$. Hence, $\psi(w) \in L$. Additionally, for $w$ and $\psi(w)$, interpreted as runs of $M$, we obtain $\eff(w) = \eff(\psi(w))$. Observe that $|\psi(w)| \leq n$ as $n$ is the length of the rejecting run of $N$ and $q_r$ has no outgoing transition from $T$.
Therefore, for each $i \in \set{1,2}$, $\eff_i(w) = \eff_i(\psi(w)) \leq n$. Therefore, the automaton recognising $ L(V_1) \cap L_2 \cap L$ needs only to simulate together $V_1$ and $V_2$, by keeping the counter values (which are between $0$ and $k+n$) in the state and verifying that we do not leave $L$ at any point using the automaton for $L$ (recall that $L$ is a regular language). Similarly, the automaton recognising the language $K\Sigma^*$ simulates $V_1$ and $V_2$ together until it encounters the first prefix $v \notin L_1 \cap L_2 \cap L$. It accepts if at least one of the following conditions holds: (i) $v \notin L_2$ or (ii) $v \notin L$ and $v \in L_1$. Otherwise, it rejects. These conditions can be checked by inspecting the last transition. Since regular languages are closed under finite union, this finishes the proof.

	\label{afterbibliography}
	\newoutputstream{pagestotal}
	\openoutputfile{main.pagestotal.ctr}{pagestotal}
	\addtostream{pagestotal}{\getpagerefnumber{afterbibliography}}
	\closeoutputstream{pagestotal}
	\newoutputstream{todos}
	\openoutputfile{main.todos.ctr}{todos}
	\addtostream{todos}{\arabic{@todonotes@numberoftodonotes}}
	\closeoutputstream{todos}
\end{document}